\documentclass[12pt,preprint]{aastex}
\usepackage{graphicx}% Include figure files
\usepackage{ulem}    % double underlines uuline
\usepackage{bm}      % bold math
\usepackage{color}
\newcommand{\rune}{{Run 1}}
\newcommand{\runea}{{Run 1} }
\newcommand{\runc}{{Run 2}}
\newcommand{\runca}{{Run 2} }
\newcommand{\runprandtla}{{Run 3-P} }
\newcommand{\runprandtl}{{Run 3-P}}
\newcommand{\runultraLa}{{Run 3} }
\newcommand{\runultraL}{{Run 3}}
\newcommand{\runhyper}{{Run 4}}
\newcommand{\runhypera}{{Run 4} }
\newcommand{\murama}{\texttt{MURaM} }
\newcommand{\muram}{\texttt{MURaM}}

\renewcommand{\vec}{\mathbf}

\def\AD#1{{\textcolor{blue}{\bf #1}}} % TEXT REVISED AFTER REVIEW
\def\neu#1{{\textcolor{blue}{\bf #1}}}  % response

\def\neu#1{#1}  %EDITORS: uncomment this line!!!
\def\AD#1{#1} % EDITORS: uncomment this line!!!

\def\resp#1{} % delete ``save''able text
\slugcomment{accepted by ApJ}

\shorttitle{Turbulent small-scale solar surface dynamo}
\shortauthors{Pietarila Graham et al.}

\begin{document}

\title{Turbulent small-scale dynamo action in solar surface simulations}

\author{Jonathan Pietarila Graham, Robert Cameron, and Manfred Sch\"ussler}
\affil{Max-Planck-Institut f\"ur 
Sonnensystemforschung, Max-Planck-Str.~2, 37191 Katlenburg-Lindau, Germany}
\email{jpietarilagraham@mailaps.org}

\begin{abstract}
We demonstrate that \AD{a magneto-convection simulation incorporating
essential physical processes governing solar surface convection}
exhibits turbulent small-scale dynamo action.  {By presenting a
  derivation of the energy balance equation and transfer
  functions for compressible magnetohydrodynamics (MHD), we
  quantify the source of magnetic energy on a scale-by-scale basis.}
We rule out the two alternative mechanisms for the generation of
small-scale magnetic field in the simulations: the tangling of
magnetic field lines associated with the turbulent cascade and
Alfv\'enization of small-scale velocity fluctuations (``turbulent
induction'').  {Instead, we find the dominant source of
  small-scale magnetic energy is stretching by inertial-range fluid
  motions of small-scale magnetic field lines against the magnetic
  tension force} to produce (against Ohmic dissipation) more
small-scale magnetic field. The scales involved become smaller with
increasing Reynolds number, {which identifies the dynamo as} a
small-scale turbulent dynamo. 
\end{abstract}


%
%________________________________________________________________

\section{Introduction}
\label{SEC:INTRO}

Generically, \neu{three-dimensional (3D)} magnetohydrodynamic turbulence excites dynamo action when
the magnetic Reynolds number exceeds a critical threshold \neu{(such that amplification by stretching dominates over Ohmic
dissipation).}  That
turbulence could amplify magnetic energy by the random stretching of
field lines was proposed by \citet{Ba1950} and first demonstrated in
direct numerical simulations by \citet{MeFrPo1981}.
\AD{Many turbulent flows with high enough
Reynolds numbers support dynamo action.  Solar surface convection is likely such a flow,}
 but the relation to the global solar dynamo is not
clear \citep{Os2003}.  There is evidence for local dynamo
action near the solar surface \citep{PeSz1993}, {and the}
question of whether surface convection can support a turbulent dynamo
is the focus of {the} present work.

\neu{The simulations of \citet{MeFrPo1981} demonstrated turbulent
  dynamos with dramatically different characteristics.  In one case,
  magnetic energy grows at scales smaller than the forcing scale of
  fluid motions.  This defines small-scale dynamo (SSD) or {\sl
    fluctuation} dynamo action (see, e.g., \citealt{IsScCo+2007}).  In
  the other case, magnetic energy grows at scales larger than the
  forcing scale: large-scale dynamo.  Such large-scale dynamos are
  often studied {in the framework of} mean-field theory (see, e.g.,
  \citealt{KrRa1980,CoTwBr+1997,FiBlCh1999,FiBl2002})} \AD{which suggests that} \neu{the production
  of large-scale magnetic energy is {related to the}
  lack of reflectional symmetry in the small-scale velocity
  fluctuations (e.g., from helical motions).  A more intuitive picture
  is that kinetic helicity creates magnetic helicity through
  Alfv\'enization, which then goes through an inverse cascade to large
  scales \citep{PFL76} -- the magnetic tension force expands coils in
  the magnetic field lines.  Symmetry breaking (e.g., helicity) is not required for
  small-scales dynamos and it is widely believed that sufficiently
  chaotic 3D flows will be SSDs.  Near the solar surface, the
  convective (granulation) time scale is much shorter than the
  rotation period, so that a flow with negligible net helicity results
  and large-scale dynamo action is not expected.}
  \AD{However, surface convection is expected to be sufficiently chaotic to be a small-scale dynamo.}

{The origin and properties of the quiet-Sun intra-network
  magnetic field provide observational evidence of and arguments for
  local small-scale dynamo action on the Sun.  Firstly,} from a
flux-transport model, \citet{PeSz1993} conclude that the decay of
active regions (or of flux tubes not quite rising to the surface) is
insufficient to account for the observed intra-network magnetic
fields.  A source term is required in their model to {match the}
observations and they conclude that the physical interpretation of
this term is small-scale dynamo action in the convection zone.
{Secondly,} small-scale dynamo action is also consistent with
high resolution magnetograms of the intra-network.  These show mixed
polarity fields on small scales (variously called the magnetic carpet
or the salt-and-pepper pattern; \citealt{TiSc1998,HaScTi2003}). The
fractal nature of {these} opposite polarities extends down to the
resolution limit of the observations \citep{PiGrDaSc2009}.
{Furthermore,} the amount of {observed} small-scale flux is not
dependent on the solar cycle, nor does it show any \neu{latitudinal}
dependence \citep{HaScTi2003,SaAl2003,TrBuShAsRa2004}.  These results
{all suggest that the source of quiet-Sun magnetic field} is
independent of the global dynamo.  {Lastly,} turbulent convection
has been shown to drive dynamo action in numerical simulations of
Boussinesq convection without rotation \citep{C99,CaEmWe2003}.
Observations and simulations together provide evidence, then, of a
small-scale dynamo driven by turbulence at the solar
surface.\footnote{Actually, simulations suggest turbulent dynamo
  action occurs in the bulk of the convection zone as well
  \citep{BMT04}.}

{An alternative interpretation of the observations is that the
  small-scale field results from turbulence acting on the
  large-scale magnetic field from the global dynamo.  Such {\sl
    induced small-scale fields} could result from two different
  processes.  One process is the turbulent tangling, sometimes called
  ``shredding,'' of field lines which moves magnetic energy to
  smaller scales as part of the turbulent energy cascade.  However,
  the turbulent cascade is associated with power-law energy spectra
  and, therefore, the amount of small-scale intra-network flux should
  change as the strength of any large-scale background field changes
  over the solar cycle.} %We can estimate the time scale of the cascade
%from the scale of the background field, $l_L$, to a smaller scale,
%$l_S$, from the Kolmogorov 1941 (hereafter, K41) phenomenology (see,
%e.g., \citealt{F95}).  The time for energy to move from scale $l$ to
%scale $l/2$ is the eddy-turnover time, $t_l\sim l/v_l$, where $v_l$
%is the velocity of turbulent eddies with diameter $l$.  At the
%granulation scale of $1\,$Mm with a typical photospheric velocity of
%$1\,$km/s, $t_{1\,{\rm Mm}}\sim10^3\,$s.  This is the time it takes
%energy to move from a scale of $1\,$Mm down to a scale of
%$500\,$km. K41 also predicts $v_l\sim l^{1/3}$ (associated with a
%-5/3 energy spectrum) which means $t_l\sim l^{2/3}$.  Therefore, the
%larger $l_L$ for the background field which is the source of the
%field at scale $l_S$, the longer it takes for changes in the
%magnitude at scale $l_L$ to propagate down to scale $l_S$. However,
%even if we take $l_L\sim1400\,$Mm, the solar diameter, changes in the
%level of back ground field should propagate to the observed
%small-scale intra-network field ($l_S\sim200\,$km) in $100\,$hours.
%This is much shorter than the 11 year solar cycle and we should then
%expect to see the amount of small-scale quiet-Sun flux vary with the
%cycle.
{This contradicts observations
  \citep{HaScTi2003,SaAl2003,TrBuShAsRa2004}. A similar argument,
  against the decay of active regions as the source of intra-network
  flux, has been put forward by \citet{SaAlEmCa2003b}.  Additionally,
  the total {unsigned magnetic flux (and energy)} in active regions even during solar
  maximum is less than the {unsigned magnetic flux (and energy)} contained in the quiet-Sun
  \citep{SA04,TrBuShAsRa2004}, making decay from active regions as a
  source of the small-scale field very unlikely.}

{The second process for producing small-scale magnetic field from
  large-scale field, called {\sl turbulent induction}
  \citep{ScIsCo+2007}, involves the stretching of a uniform (or large
  scale) background magnetic field by turbulent fluid motions which
  excites Alfv\'enic magnetic fluctuations on the same scale as the
  turbulent fluid eddies.  Alfv\'enic turbulent induction will be
  present whenever there is a significant background field. In the
  presence of both a flow capable of sustaining small-scale dynamo
  action and a large-scale magnetic field,\footnote{{For very strong background fields (several
    times the equipartition field strength, $B_{\rm
      eq}\approx\sqrt{4\pi\rho_0}v_{rms}\approx350\,$G), the
    large-scale magnetic field quenches the dynamo
    \citep{HaBr2004b}.}} the small-scale dynamo and
  Alfv\'enic induction may compete as the source of small-scale
  field.}

Given the prevalence of turbulent dynamos in homogeneous, isotropic
turbulence, {such dynamo action is expected in the Sun unless
  additional physics can be identified which would suppress it. 
Two points are often raised to argue that a small-scale dynamo cannot
operate in the Sun.}
Firstly, turbulent eddies smaller than the characteristic scale of the
magnetic field act like a turbulent magnetic diffusivity and {could} inhibit dynamo
action.  This is an important concern for the Sun (and other stars)
because the kinetic Reynolds number, $Re \equiv v_0l_0/\nu$ ($v_0$ and
$l_0$ {being} typical velocity and length scales, and $\nu$ the
kinematic viscosity), is much larger than the magnetic Reynolds
number, $Re_M \equiv v_0l_0/\eta$ ($\eta$ {being} the magnetic
diffusivity).  Their ratio, the magnetic Prandtl number, $P_M \equiv
Re_M/Re$, is approximately $10^{-5}$ near the solar surface.
%{Kolmogorov 1941 (hereafter, K41) phenomenology
%  predicts the scales of the smallest eddies to be $l_\nu\sim
%  Re^{-3/4}$ (see, e.g., \citealt{F95}).} This means that turbulent
%eddies exist on scales down to roughly $1/6000$ the size of the
%smallest magnetic fluctuations.
The critical magnetic Reynolds number
for turbulent dynamo action, $Re_M^C$, sharply increases with
decreasing $P_M$ {and it has been suggested that $Re_M^C$ goes to infinity as $P_M$ goes to zero} \citep{RoKl1997,BoCa2004,ScHaBr+2005}. %and two asymptotic possibilities exist.  Either
%$Re_M^C$ approaches a constant as $Re$ approaches infinity
%(small-scale dynamo at low $P_M$; \citealt{RoKl1997,BoCa2004}) or
%their ratio $Re_M^C/Re=P_M^C$ approaches a constant (no small-scale
%dynamo at low $P_M$; \citealt{ScHaBr+2005}).
{However, recent numerical simulations \AD{attest that $Re_M^C$
approaches a plateau for $P_M\ll1$} \neu{both with physical (Laplacian) viscosity \citep{PMM+05}
and with hyperviscosity
\citep{IsScCo+2007}.
Baring the identification of a new length scale in the problem, these results
indicate that, for magnetohydrodynamics (MHD), small-scale dynamo action remains possible in
the asymptotic limit as $P_M$ goes to zero.
Additionally,
a laboratory dynamo with liquid sodium
($P_M\approx10^{-5}$) resulting from unconstrained turbulence has been
demonstrated \citep{MoBeBo+2007}.  This establishes that a turbulent
(at least, a large-scale) dynamo is possible at values of $P_M$ corresponding to the solar
plasma  \citep{MoBeAu+2009}.}

{The second suppression argument is that,} unlike the \citet{C99}
Boussinesq simulation, the Sun is strongly stratified and magnetic
flux is swept into the downflows and subject to long recirculation
times \citep{StBeNo2003}.  {In their stratified simulations with
  open boundaries, \citet{StBeNo2003} found no evidence of dynamo
  action for surface convection.  Their magnetic Reynolds number,
  however, was below the critical value.  \citet{VoSc2007} have
  demonstrated that surface convection with little field recirculation
  and strong density stratification (as well as other physical effects
  present in the Sun) can support local dynamo action when
  $Re_M\ga2000$.} \AD{We will determine if this local dynamo action is
a small-scale turbulent dynamo.  Currently,}
 no other likely suppression mechanisms for
  small-scale dynamo action in the intra-network photosphere are
  known.

The magnetic energy spectrum {of the \citet{VoSc2007} dynamo}
peaks at scales smaller than the energy-containing scale of the fluid
motions, which is suggestive of {a} small-scale dynamo.
{However, {except in the most idealized of simulations,\footnote{The case of delta-correlated
    in time, isotropic, and homogeneous forcing being the sole
    exception.}} non-zero
  time-averaged mean flows exist for times ($\sim10\,$minutes for photospheric
  convection) long compared to inertial-range eddy-turnover times
  (e.g., $\sim6\,$seconds at a scale of $1\,$km).  Therefore, even in
  the absence of net helicity, this mean flow can act as a low $Re_M$
  dynamo that produces magnetic field near the mean-flow scale
  \citep{PoMiPi+2007}, $\approx1\,$Mm for convective granulation.
  This would not be a small-scale dynamo; it would \AD{still have} small-scale magnetic field \AD{produced} from either the
  turbulent cascade or from Alfv\'enic turbulent induction.}  \AD{In order to fully understand what is occurring,} {it is
  important to disentangle the possible sources of small-scale
  magnetic energy} \AD{and} properly identify the dynamo mechanism
  \citep{ScIsCo+2007}.

We have employed transfer analysis to measure, scale by scale,
the relative strengths of the sources of magnetic energy: turbulent
cascade, {Alfv\'enic turbulent induction,} and dynamo {action (by stretching of
field lines)} in the
{simulations of} \citet{VoSc2007}. The scales involved in the dynamo
and their dependence on Reynolds number as well as growth rates and
energy spectra allow us to determine {that the} dynamo is a turbulent
small-scale dynamo.  %These measurements also allow us to quantify the
%difference and similarities with the small-scale dynamos of
%homogeneous, isotropic, and incompressible turbulence.

%__________________________________________________________________

\section{{Data and methods}}

We \AD{use \murama \citep{V03,VSS+05} to perform simulations} for a rectangular
domain of horizontal extent $4.86 \times 4.86\,$Mm$^2$ and a depth of
$1.4\,$Mm ($800\,$km below and $600\,$km above the {simulated solar}
surface).  \neu{The sides boundaries
  are periodic in both horizontal directions.  The open lower boundary
  assumes upflows to be vertical, $v_x = v_y = \partial_zv_z=0$; for
  downflows vertical gradients are set to zero, $\partial_zv_x =
  \partial_zv_y = \partial_zv_z=0$.  The upper boundary is closed.}
  \neu{The magnetic field is vertical at upper and lower boundaries: $B_x =
  B_y = \partial_zB_z=0$.  This ensures that there is no Poynting flux
  into or out of the box.  The magnetic diffusivity is increased in
  the lower $150\,$km of the box in order to well resolve the
  diffusive boundary layer \citep{VoSc2007}.
\AD{These boundary conditions allow the convective downflows to leave the
box and, thus, simulate an artificially isolated surface layer.  These
same downflows drag magnetic field to the lower region of enhanced
diffusivity where it can be eliminated (this simulates the fact that the
field should not be available for further stretching due to long recirculation times; see \citealt{StBeNo2003}).}
 The upper $600\,$km of
  the box is convectively stable and the lower $800\,$km of the box is
  convectively unstable.  The convection is driven by radiative
  cooling} \AD{(calculated using grey radiative transfer and the Rosseland mean opacities)} \neu{at the surface where the optical depth is unity.}
\AD{The effects of partial ionization are included in the equation of state.}  \neu{These
  simulations are then so-called ``realistic'' simulations of a
   portion of the solar surface layer} \AD{including compressibility and stratification (4 orders-of-magnitude variation in the density), radiative energy transport, partial ionization, and little recirculation.}   \AD{``Realistic'' is used
here} \neu{to distinguish them from a class of
  ``idealistic'' simulations of incompressible, isotropic,
  homogeneous, and triply-periodic MHD turbulence.  Of course, neither
  type of simulation is able to achieve the Reynolds and Prandtl
  numbers of the Sun, but aside from this, \murama simulations} \AD{include
more of the physics relevant to the near-surface layers of} the Sun.
 Our aim is to determine if
these effects inhibit turbulent small-scale dynamo action or
overshadow it with some other mechanism.

%\label{TABLE:RUNS}
\begin{deluxetable}{lccccc}
\tablewidth{0pt} 
\tablecaption{Properties of \murama \neu{dynamo} runs: grid
  resolution, \neu{effective magnetic Prandtl number, $P_{M,{\rm eff}}$,} magnetic diffusivity, $\eta$, magnetic Reynolds number,
  $Re_M$ based on $v_{rms}\approx3.3\,$km/s and a length scale of
  $1\,$Mm, and kinematic \neu{dynamo} regime $e-$folding time, $\gamma^{-1}$.}
\tablehead{ \colhead{Simulation} &
  \colhead{Grid Resolution (km)} &\colhead{\neu{$P_{M,{\rm eff}}$}} & \colhead{$\eta$ (cm$^2$ s$^{-1}$)} &
  \colhead{$Re_M$} & \colhead{$\gamma^{-1}$ (s)} }
\startdata
% l_v = 161.24, l_m = 65.19, P_M = c * (l_v/l_M)^2 = 1/3*6.12 
\runea &  $9\times9\times10$ &  \neu{$\sim2.0$} & $1.6\cdot10^{10}$ & $\approx 2100$&  $1600$ \\ %\hline
% l_v = 120.81, l_m = 61.31, P_M = c * (l_v/l_M)^2 = 1/3*3.88 
\runca &  $7.5\times7.5\times10$ & \neu{$\sim1.3$} &  $1.25\cdot10^{10}$ & $\approx 2600$ &  $660$ \\ %\hline
% l_v = 78.69, l_m = 51.39, P_M = c * (l_v/l_M)^2 = 1/3*2.34
\runprandtl  & $5\times5\times7$ &  \neu{$\sim0.8$}  &  $1.25\cdot10^{10}$ & $\approx 2600$ &  $1200$\\ %\hline
% l_v = 81.86, l_m = 44.58, P_M = c * (l_v/l_M)^2 = 1/3*3.37 
\runultraLa &  $5\times5\times7$ & \neu{$\sim1.1$} &   $6.25\cdot10^{9}$ & $\approx 5300$ &  $230$ \\ %\hline
% l_K = 65.95, l_m = 35.86, P_M = c * (l_v/l_M)^2 = 1/3*3.38
\runhypera &  $4\times4\times4$ & \neu{$\sim1.1$} &  $4\cdot10^{9}$ & $\approx 8300$ &  $150$ \\ %\hline
\enddata
%%%\tablenotetext{a}{Star}
%%% You can append references to a table using the \tablerefs command.
\label{TABLE:RUNS}
\end{deluxetable}

\neu{Dynamo runs} with increasing resolution
and Reynolds numbers have been carried out.  Table \ref{TABLE:RUNS}
summarizes the parameters of the runs.  \neu{Figure \ref{FIG:EMVST} displays the time
evolution of magnetic energies.} \runca has previously been
reported (as ``Run C'') for dynamo action in \cite{VoSc2007}.  {All runs} start
from an initial hydrodynamic case plus weak magnetic seed field.
There is an initial growth of $E_M$ for about $3\,$minutes due to flux
expulsion and convective intensification followed by the linear
(kinematic) phase during which magnetic energy grows exponentially
  with time and the magnetic field is too
  weak to affect fluid motions. The lower resolution simulations are
  continued until the nonlinear phase {when} the back reaction of the
  Lorentz force on fluid motions becomes important and the magnetic
  energy begins to saturate.

%    \label{FIG:EMVST}

\begin{figure}[htbp]
  \plotone{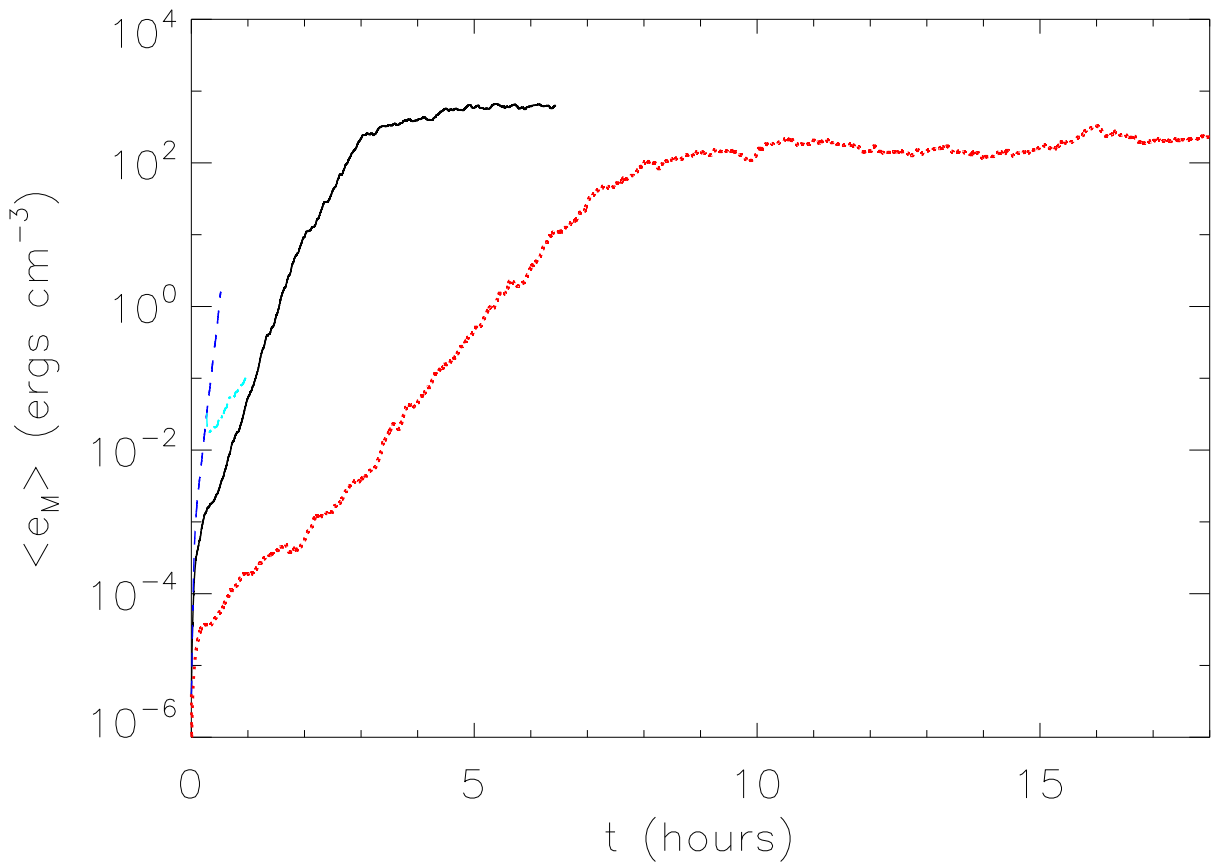}%{.eps}
  \caption{Mean magnetic energy density, $\langle e_M \rangle$, versus
    time, $t$, for dynamo simulations.  Shown are \runea (red,
    dotted), \runca (black,
    solid), \runultraLa (blue, dashed), and \runprandtla (cyan, dash-dotted). }
    \label{FIG:EMVST}
\end{figure}

\neu{For {all \murama} simulations, constant magnetic diffusivity is
  employed (outside the region of enhanced diffusivity near the bottom
  boundary), but artificial shock resolving and hyperviscosity are
  applied to the momentum equations (see \citealt{VSS+05} for
  details).}  \AD{This approach allows us to reach high
effective kinetic Reynolds number, $Re_{\rm eff}$, without a prohibitively
small time step.  No value of viscosity is defined, but $Re_{\rm eff}$
   can be estimated
  from the energy spectra (Figure \ref{FIG:SCALES}). The estimate of the
effective Reynolds number is
 derived from the representative length scales:}
  the velocity Taylor microscale, $\lambda_V$,
\begin{eqnarray}
\lambda_V^2 = \langle v^2 \rangle \big{/} \langle \omega^2 \rangle
= \int_0^\infty E_V(k_\perp) dk_\perp \big{/}\int_0^\infty k_\perp^2E_V(k_\perp) dk_\perp
\end{eqnarray}
where $E_V(k_\perp)$ is the horizontal velocity spectrum\footnote{We employ horizontal spectra which
    are obtained by performing a two-dimensional Fourier transform of
    any field, $f$, at each horizontal layer resulting in
    $\hat{f}(k_x,k_y,z,t)$.  This quantity is then projected onto a
    one-dimensional wavenumber $k_\perp^2=k_x^2+k_y^2$.  The
    horizontal spatial frequency, $\nu_\perp$, is $k_\perp/2\pi$.} and
 the integral scale for the turbulent motions, $L_0$, 
\begin{equation}
L_0 = \int_0^\infty k_\perp^{-1}E_V(k_\perp) dk_\perp \big{/}\int_0^\infty E_V(k_\perp) dk_\perp\,.
\label{eq:L0}
\end{equation}
The
 effective Reynolds numbers
of the simulations is given by \citep{Ba1970,WeMaDa+2007},
\begin{equation}
Re_{\rm eff} \propto \bigg{(} \frac{L_0}{\lambda_V}\bigg{)}^2\,,
\label{eq:reeff}
\end{equation}
\neu{where the constant of proportionality is unknown.
We can also determine the magnetic energy Taylor
scale, $\lambda_M$,
\begin{eqnarray}
\lambda_M^2 
= \int_0^\infty E_M(k_\perp) dk_\perp \big{/}\int_0^\infty k_\perp^2E_M(k_\perp) dk_\perp\,.
\end{eqnarray}
This allows us to estimate the effective magnetic Prandtl number,
\begin{equation}
P_{M,{\rm eff}} \propto \bigg{(} \frac{\lambda_V}{\lambda_M}\bigg{)}^2\,.
\label{eq:pmeff}
\end{equation}
We determine the constant of proportionality from an incompressible,
isotropic, and homogeneous} \AD{dynamo simulation.  We employed
a pseudospectral code \citep{GMD05b,GMD05} and forced the velocity field
with a superposition of harmonic modes with random phases,
the resolution was $256^3$ grid points, and $P_M=4$.
From this calibration of Eq. (\ref{eq:pmeff}),
we estimated that the magnetic Prandtl numbers in our \murama runs are
between 1 and 2 (see Table \ref{TABLE:RUNS}).}

% \label{FIG:SCALES}
\begin{figure}[htbp]
  \plottwo{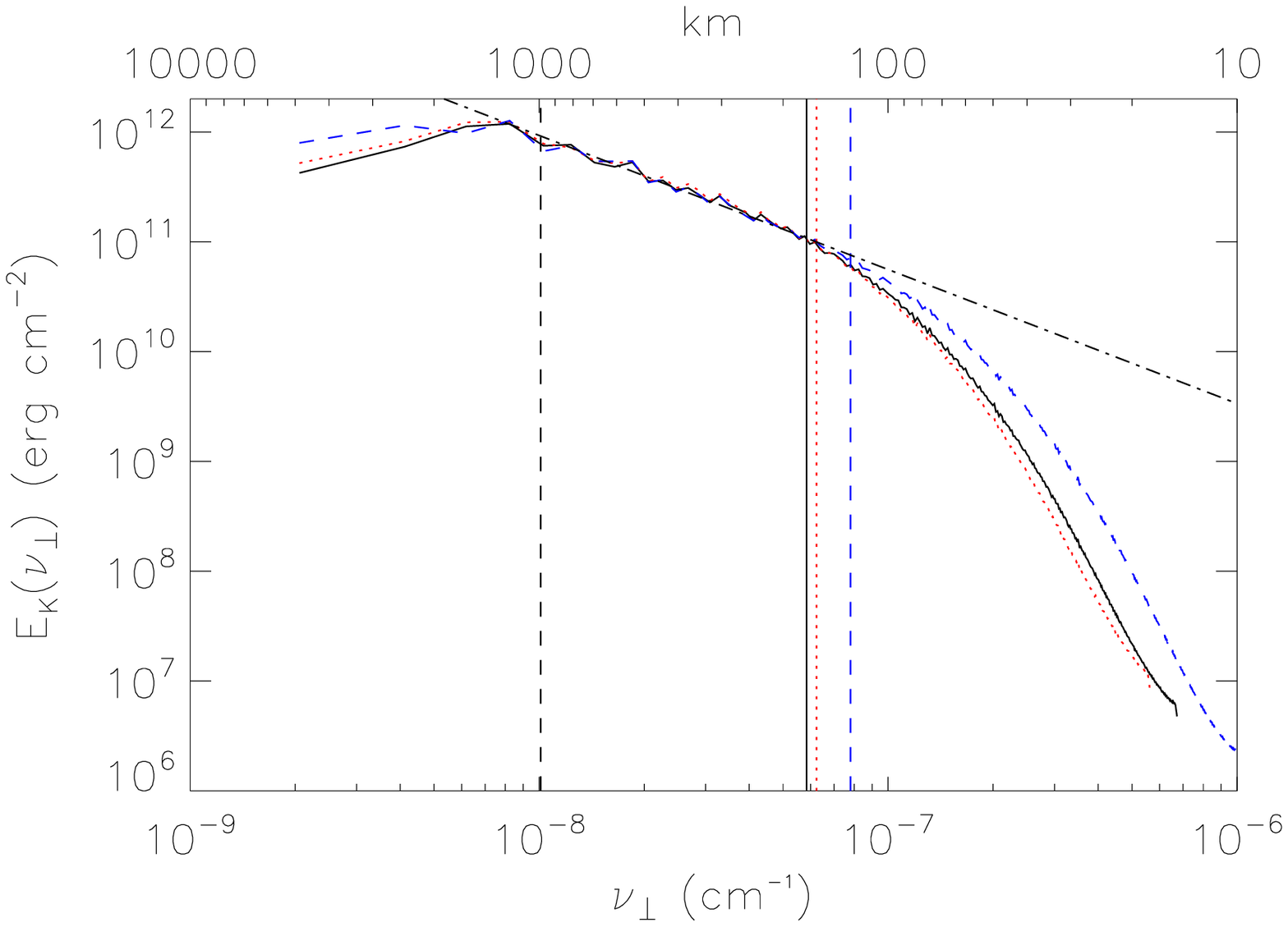}{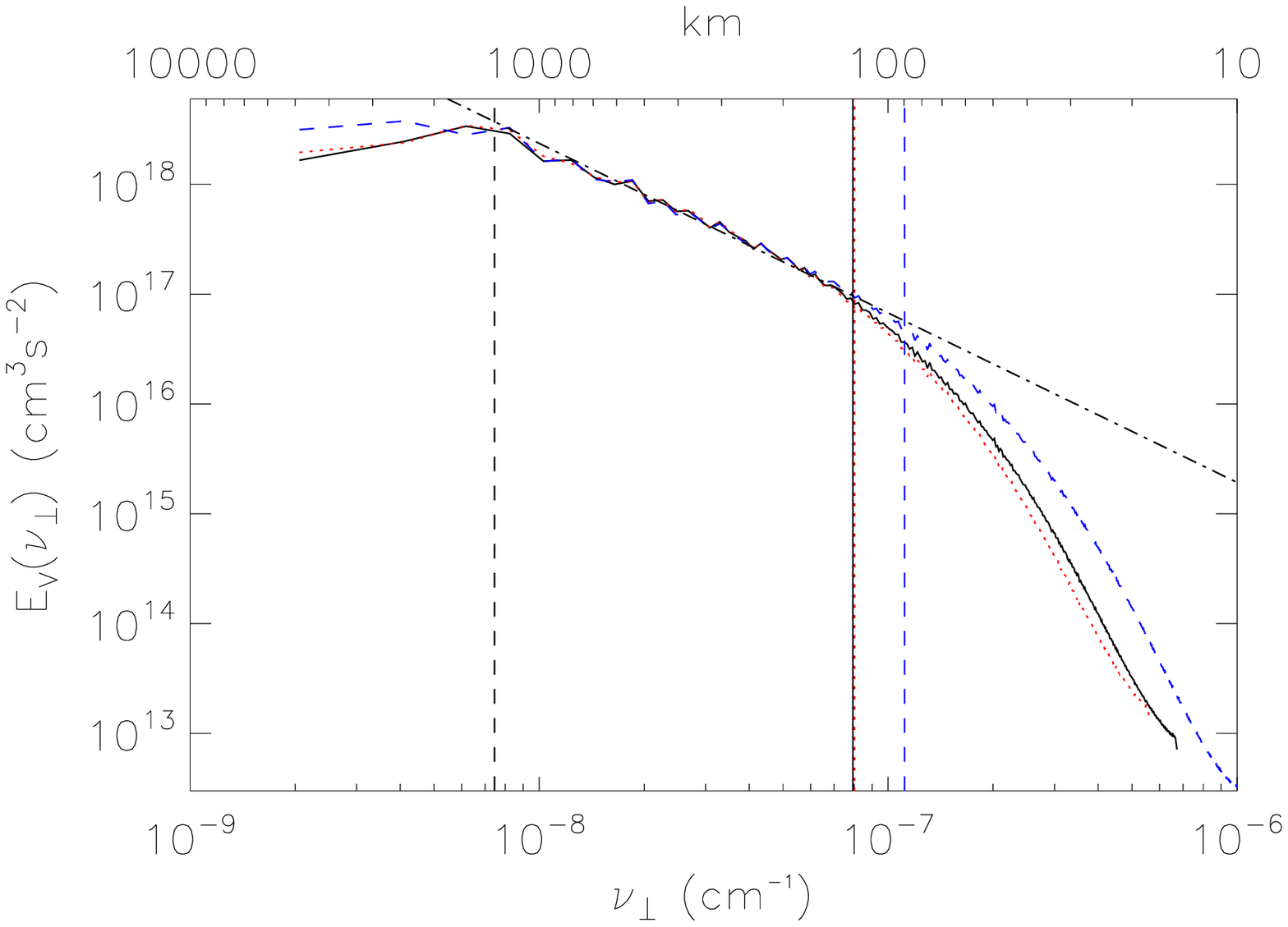}
  \caption{Left: Kinetic energy spectrum, $E_K$, versus horizontal
    spatial frequency, $\nu_\perp$, for \runca (black,
    solid) averaged over
    $t\in[2.2,2.4]\,$hr, for \runea (red, dotted) averaged over $t\in[0.2,1.4]\,$hr,
    and for \runultraLa (blue, dashed) averaged over $t\in[0.1,0.5]\,$hr.  The vertical lines represent the
    integral scale ($\approx10^8\,$cm) and Taylor scales {near}
    the beginning and end of the inertial range where we identify a
    power-law scaling: $k_\perp^{-1.2}$ (dot-dashed line). Right:
    Velocity spectrum, $E_V$, versus $\nu_\perp$ for the same {runs}.  The
    identified power law extends to smaller scales and has a slope of
    $-1.54\pm0.07$.}
  \label{FIG:SCALES}
\end{figure}

\neu{We find power-law scalings, $k_\perp^{\beta}$, in the inertial
  range with $\beta=-1.2\pm0.1$ for the kinetic energy spectrum and
  $\beta=-1.54\pm0.07$ for the velocity spectrum.  Using second-order
  longitudinal structure functions for the velocity (not shown), we
  determined that at scales larger than $\approx50\,$km (approximately
  half a pressure scale height} \AD{of the vertical stratification} \neu{at the surface) the flow is
  anisotropic.  The strong vertical downflows in the convectively
  unstable lower $800\,$km of the box are the source of this anisotropy. Because of the anisotropy, our spectra cannot be compared to the
  $\beta=-5/3$ from the Kolmogorov spectrum for homogeneous, isotropic turbulence (see, e.g.,\citealt{F95}). }

%__________________________________________________________________
\section{\neu{Comparison with ``idealistic'' small-scale dynamos}}

\subsection{Incompressible, isotropic kinematic/linear phase}

In the kinematic regime, where the Lorentz force is insignificant, the
magnetic energy is observed to grow exponentially, $E_M \sim \exp
\gamma t$ ($\gamma$ is the growth rate).  For turbulent dynamos (with $P_M
\ll 1$), {the growth rate} scales as
\begin{eqnarray}
\gamma \sim  Re_M^{1/2}\,,
\label{eq:gamma2}
\end{eqnarray}
{though no simulation has yet exceeded} $Re_M^C$ enough to observe the
predicted scaling (see, e.g., \citealt{ScIsCo+2007}).  For the large
magnetic Prandtl case, $P_M \ge 1$, we expect $\gamma\sim Re^{1/2}$.
Again, this scaling has {not yet been found in any simulations}
\citep{ScCoTa+2004}.  \neu{Other predictions}
about the kinematic (linear) phase of the
small-scale dynamo \neu{are} due to the exactly solvable Kazantsev model (see
\citealt{ScCoTa+2004} and references therein).  This model predicts
that the amplitude of each (Fourier) mode grows exponentially at the
same rate (see also, \citealt{MPM+05} for a growth rate analysis) and
that the magnetic energy spectrum at large scales follows a $k^{3/2}$
scaling.  

\subsection{\murama kinematic/linear phase}
\label{SEC:LINEAR}

  For the \murama
dynamo, the dependence of growth rate on Reynolds number is shown in
Figure \ref{FIG:GAMMA}.  {The growth rate is an increasing
  function of Reynolds number.  This indicates that the dynamo is a
  small-scale (inertial range) process.}  We find that we have not yet
sufficiently exceeded $Re_M^C$ \neu{(from \citealt{VoSc2007} and \rune,
$1300<Re_M^C<2100$)} to observe {the predicted}
power-law scaling. 
 {Instead, at the Reynolds numbers we can
  achieve, the growth rate {depends almost linearly on $Re_M$} (dashed
  line).}  Higher resolution {simulations} will be required to determine if there
is, indeed, an asymptotic power-law.

%  \label{FIG:GAMMA}
\begin{figure}[htbp]
  \plotone{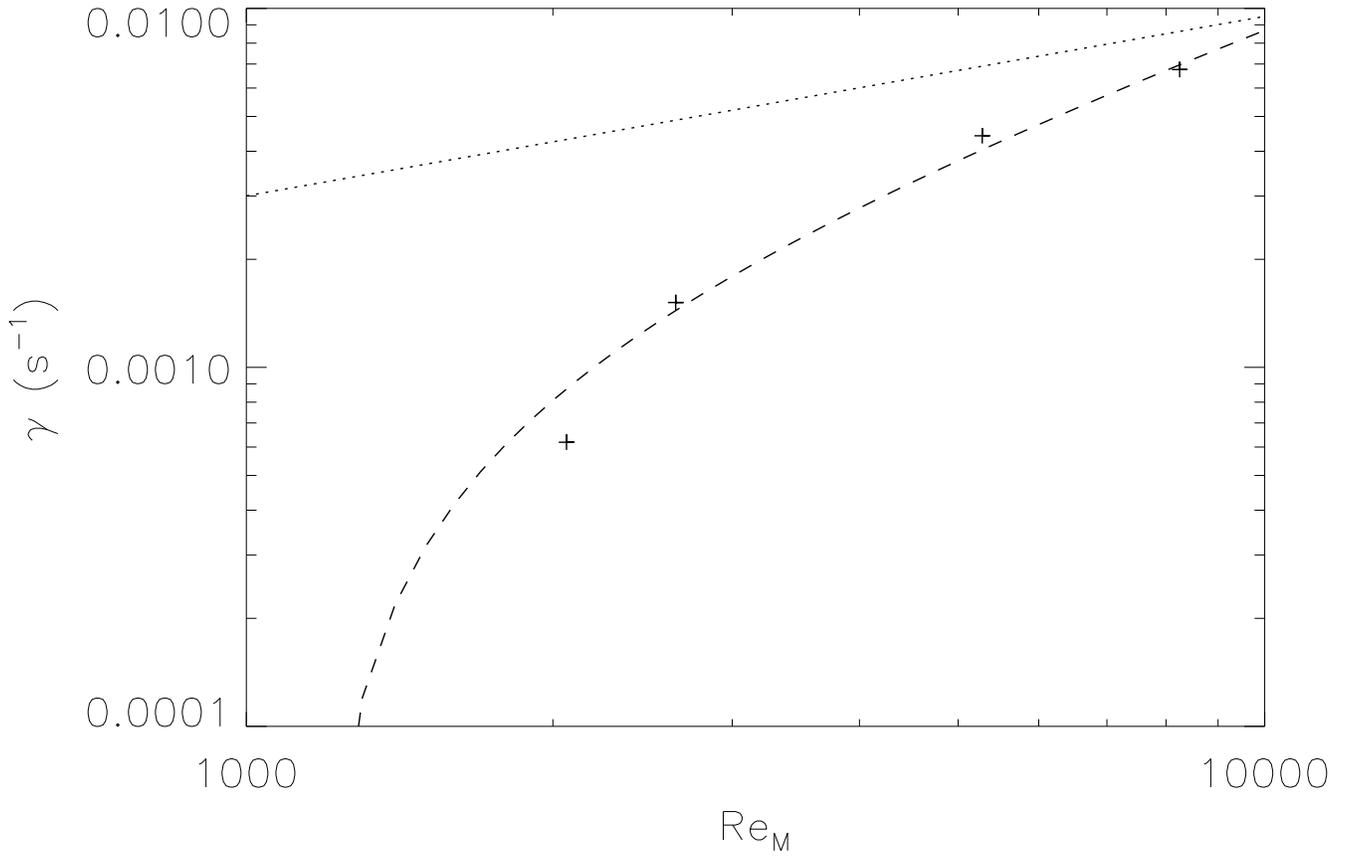}
  \caption{Growth rate {of magnetic energy,} $\gamma$, versus
    magnetic Reynolds number, $Re_M$ for \rune, \runc, \runultraL, and
    \runhyper.  Also shown is a linear fit (dashed) and the
    theoretical $Re_M^{1/2}$ scaling (dotted) for reference.}
  \label{FIG:GAMMA}
\end{figure}

In Figure \ref{FIG:LINGROWTH}, we have plotted magnetic energy
spectra, $E_M(\nu_\perp)$, at different times from the beginning of
the linear regime (including the end of the flux expulsion phase) for
\runultraL.    Power
laws can be fit to the magnetic spectrum for scales larger than
$\lambda_K$, where magnetic energy is growing at the same rate for all
scales.  A power law is in contrast to the case of a laminar dynamo
and indicates {that} a turbulent, self-similar process underlies the
dynamo mechanism.  We find $E_M(k)\propto k_\perp^\beta$ with
$\beta=0.83\pm0.04$.  The results from \runca are $\beta=0.63\pm0.06$
and for \rune, $\beta=0.53\pm0.05$.  None of these slopes agree with
$k^{3/2}$ from the isotropic Kazantsev case, but are not expected to
because of the anisotropy of fluid motions in the downflows.  \neu{We
  find, however, that the degree of anisotropy decreases with
  increasing $Re_{\rm eff}$ for the runs and the exponent $\beta$
  becomes closer to $3/2$.  This decrease in anisotropy} is due to the
generation of stronger horizontal fluctuations by the turbulence
(strong vertical fluctuations are induced by the convective driving).
In summary, we find that the general character of the \murama dynamo
is similar to the \neu{incompressible, isotropic, small-scale} dynamo \neu{but} the
spectral index differs.

%  \label{FIG:LINGROWTH}
\begin{figure}[htbp]
  \plotone{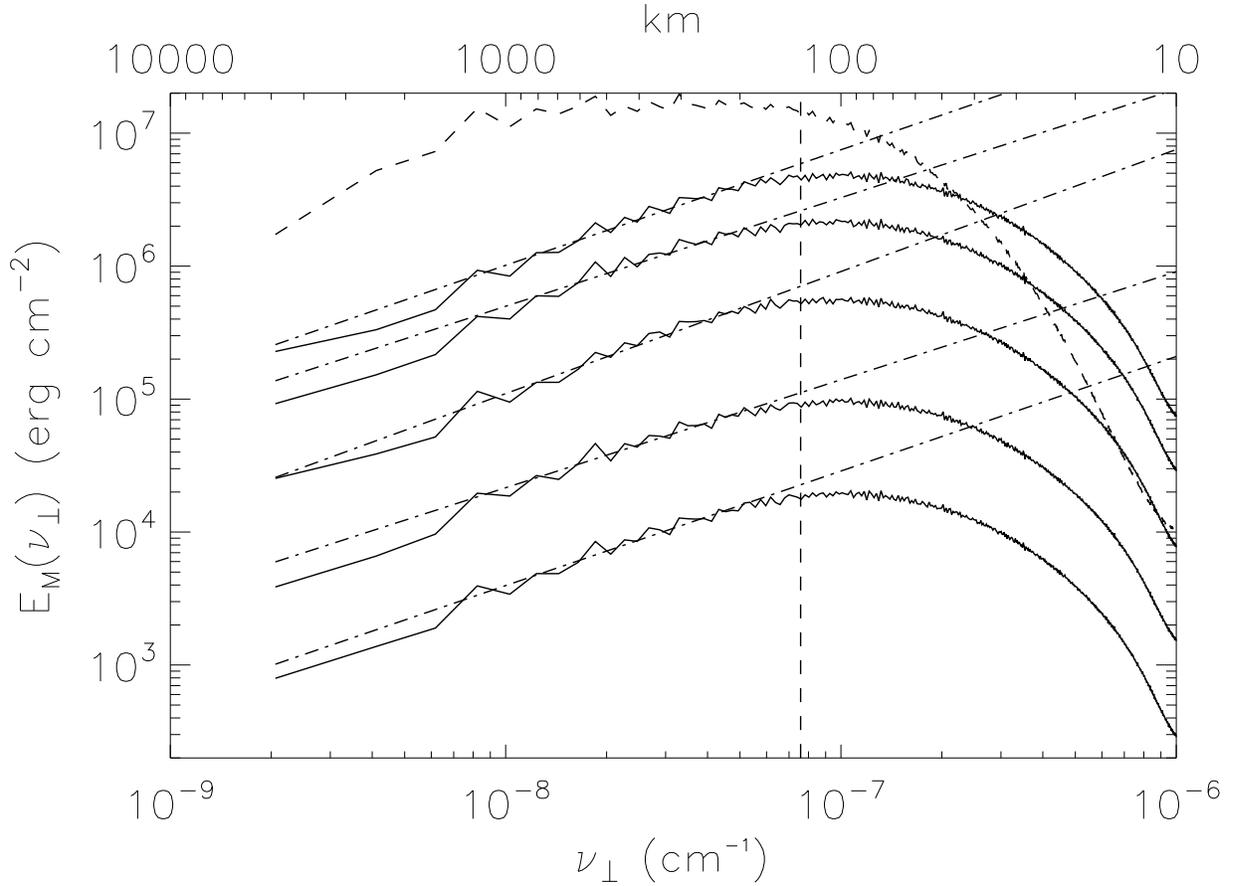}
  \caption{Magnetic energy spectra, $E_M(\nu_\perp)$ (solid lines),
    for \runultraLa for $t=11$, $16$, $23$, $ 28$, and $31\,$min (kinematic phase).  The
    vertical dashed line is the kinetic Taylor scale.  Above this
    scale, the magnetic energy spectra follow a power law (indicating
    a turbulent dynamo process) with all scales growing at the same
    temporal rate.  The time-averaged kinetic energy spectrum (dashed
    line) is compensated by $k_\perp^{1.2}$ so the inertial range is
    flat.}
  \label{FIG:LINGROWTH}
\end{figure}

\section{Transfer analysis {(kinematic phase)}}
\label{SEC:TRANSFER}

It is clear that the magnetic energy is peaking at scales nearly one
order of magnitude smaller than the energy-containing scale of the
fluid motions, $L_0\approx1\,$Mm (see Figure \ref{FIG:LINGROWTH}).  This {indicates} small-scale dynamo action,
with one caveat.  {As discussed in Section \ref{SEC:INTRO},
  {``mean flows''} might give rise to a Mm-scale dynamo.  We must,
  therefore,} disentangle three possible sources of small-scale
magnetic energy: the tangling of field lines caused by the turbulent
energy cascade,
Alfv\'enic response of larger-scale field to smaller scale velocity
fluctuations (Alfv\'enic turbulent induction), and dynamo
stretching of magnetic field lines \citep{ScIsCo+2007}.  This can be
accomplished by spectral transfer analysis.

{Spectral transfer analysis was introduced by \citet{Kr1967} and
  is widely used to understand \neu{incompressible} turbulent processes in both two (e.g.,
  \citealt{MaVa1993,Ey2006}) and three (e.g., \citealt{Zh1993})
  dimensions for both Navier-Stokes (e.g.,
  \citealt{Kr1971,MiAlPo2008}) and MHD
  (e.g., \citealt{DeVeCa2005,VeAyCh2005}).  In Appendix
  \ref{SEC:TRANSFER}, we extend it to compressible
  MHD.  Transfer analysis allows us not only to quantify the sources
  of magnetic energy but also the scales at which they operate and the
  scales at which they generate magnetic energy.  In general, the
  transfer function $T_{XYF}(k)$ measures the net rate of energy
  transfer from energy reservoir $X$ to reservoir $Y$ mediated by
  force $F$.  For $T_{XYF}(k) >0$ ($<0$) net energy is received (lost)
  by $Y$ at the scale given by $2\pi/k$.  Transfers between the
  kinetic and magnetic energy reservoirs occur via the Lorentz force,
  which can be separated into the effects of a magnetic pressure and a
  magnetic tension force.  For example, $T_{BKT}(k)$ measures the net
  work done by the magnetic tension force on the fluid motions at
  wavenumber $k$ by {\sl all scales} of the magnetic field.  When it
  is negative, net kinetic energy is lost by fluid motions at
  wavenumber $k$ working against the magnetic tension force.
  Likewise, $T_{KBT}(k)$ measures the net work done on the magnetic
  field at wavenumber $k$ by all scales of fluid motions.  These two
  transfer functions measure energy exchanged by the kinetic and
  magnetic energy reservoirs and, therefore, the rate of energy gained
  by the magnetic field is equal to the negative of the rate of energy
  lost by fluid motions,
\begin{equation}
\sum_kT_{KBT}(k)=-\sum_kT_{BKT}(k)\,.
\end{equation}
Similarly,
  $-T_{BKP}(k)$ measures work against magnetic pressure gradients and
    $T_{KBP}(k)$ the net magnetic energy generated, with
\begin{equation}
\sum_kT_{KBP}(k)=-\sum_kT_{BKP}(k)\,.
\end{equation}
This second transfer is peculiar to compressible MHD as $\sum_kT_{KBP}(k)\equiv0$
for the incompressible case \neu{where} $T_{KBP}$ measures only the effects of
the turbulent cascade moving magnetic energy to smaller scales.
In our case $T_{KBP}$ measures both the magnetic portion of the turbulent
cascade and magnetic energy generated by compressive motions.}

%  \label{FIG:TRANSFERS}
\begin{figure}[htbp]
  \plotone{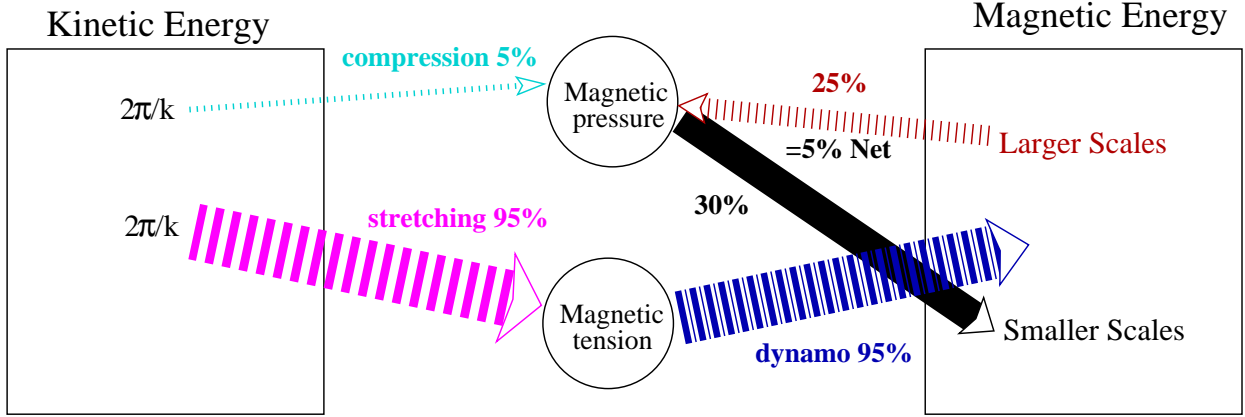}
  \caption{Net energy transfers in the kinematic phase.  Work against
    magnetic tension, $T_{BKT}<0$ (stretching), accounts for
    {95\%} of magnetic energy generated while work against magnetic
    pressure, $T_{BKP}<0$ (compression), accounts for {5\%}.  The
    latter is involved in the process of breaking down larger-scale
    field, $T_{KBP}<0$ (25\%), into smaller-scale field,
    $T_{KBP}>0$ (30\%), as part of the turbulent cascade.  The
    dominant producer of magnetic energy is the stretching of magnetic
    field lines against the magnetic tension force, $T_{KBT}>0$ (associated with turbulent dynamo action).}
  \label{FIG:TRANSFERS}
\end{figure}

{In Figure \ref{FIG:TRANSFERS}, we illustrate the net energy
  transfers between kinetic and magnetic energies, and in Figure
  \ref{FIG:RUNC_LIN} we plot their scale dependence.  We find that the
  dominant source ({95\%}) of magnetic energy is the
  stretching of field lines against the magnetic tension force.
  A lesser source of magnetic energy
  ({5\%}) is work against the magnetic pressure force.  This
  latter energy is combined with larger-scale ($>30\,$km) magnetic
  energy lost ({25\%}) to the ``magnetic cascade'' to
  produce ({30\%}) smaller-scale ($<30\,$km) magnetic
  energy.}  Using a similar expression to Equation (\ref{eq:L0}), we
determine that the predominant scale at which fluid motions are doing
work against the {magnetic tension} force {to stretch magnetic
  field lines} is {$140\,$km.}  As seen in Figure
\ref{FIG:RUNC_LIN} for \runultraL, $-T_{BKT}$ peaks at the
corresponding spatial frequency, {$7\cdot10^{-8}\,$cm$^{-1}$.}
{Acceleration of large-scale fluid motions by the magnetic tension
force, $T_{BKT}>0$ green dash-dotted line, shown in Figure
\ref{FIG:RUNC_LIN} is mostly likely due to the somewhat artificial
separation of the Lorentz force into magnetic tension and an isotropic
magnetic pressure. As there is no Lorentz force along magnetic field
lines, there can be no transfer along field lines and a portion of
negative $T_{BKP}$ is offset by positive $T_{BKT}$.} To see at which
scales the magnetic field is gaining or losing energy, we examine
$T_{KBT}$.  We find that net magnetic energy is gained at
all scales of the simulation.  The predominant scale for magnetic
energy production {by stretching} is identified as {$65\,$km}
(a spatial frequency of {$1.5\cdot10^{-7}\,$cm$^{-1}$).}
{Work against magnetic pressure, $T_{BKP}<0$ shown as a cyan
  dotted line, is mainly by granulation-scale fluid motions.  The net
  result is to remove ($T_{KBP}<0$, red dotted line) larger-scale
  magnetic energy, which was produced by stretching motions, and break
  it down into smaller-scale magnetic structures ($T_{KBP}>0$, black
  solid line). }

%  \label{FIG:RUNC_LIN}
\begin{figure}[htbp]
  \plottwo{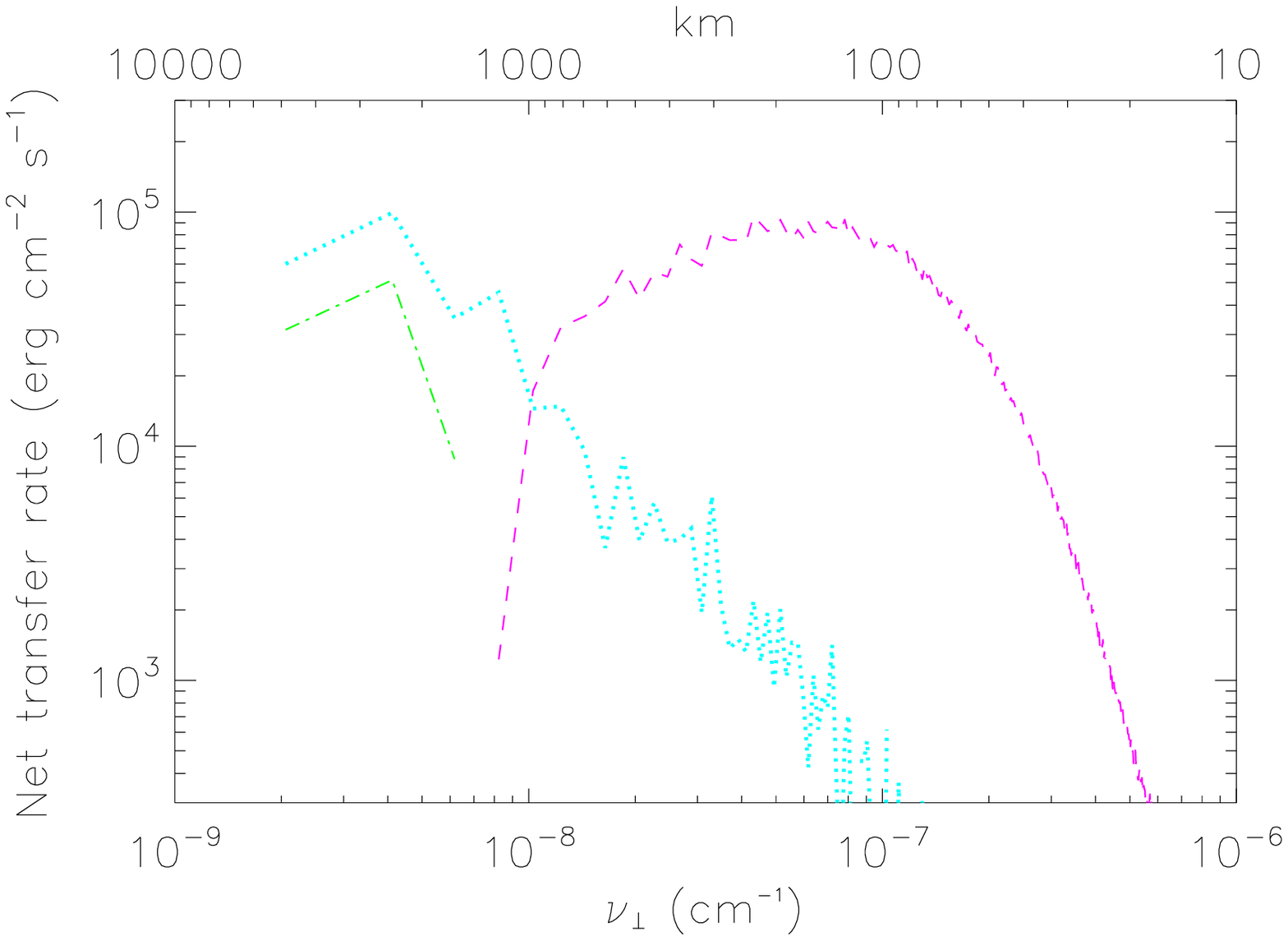}{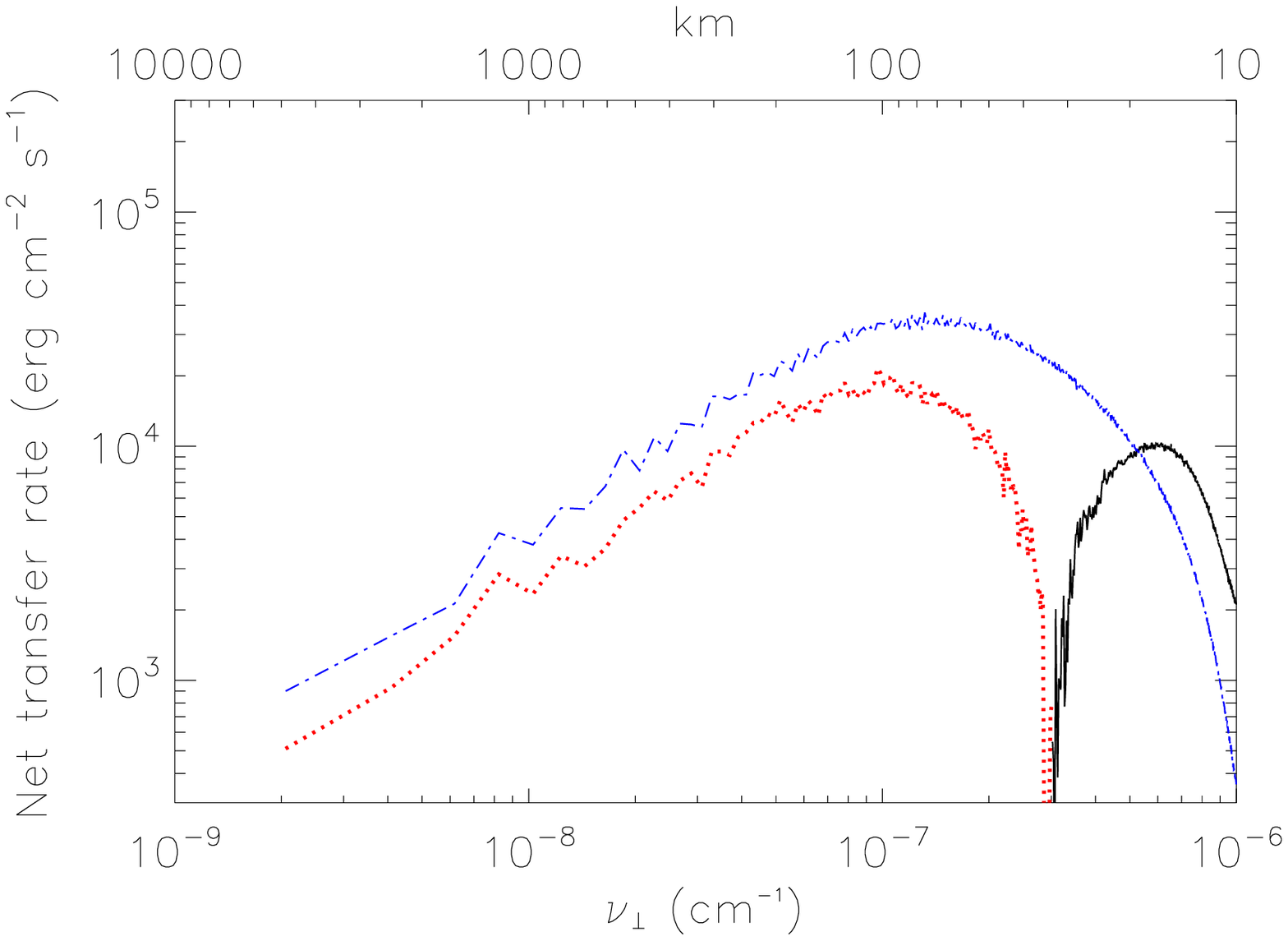}
  \caption{Net energy transfer rates for \runultraLa versus horizontal
    spatial frequency, $\nu_\perp$, averaged over $t\in[11,28]\,$min
    (linear dynamo regime). {Left:} Work against {magnetic
      tension, $T_{BKT}(k)<0$, is shown as pink dashed lines; fluid
      acceleration by magnetic tension, $T_{BKT}(k)>0$, as green
      dot-dashed; and work against magnetic pressure, $T_{BKP}(k)<0$,
      as cyan dotted.  {Right:} Dynamo stretching, $T_{KBT}>0$, is
      shown as blue dot-dashed lines; generation by compression (and
      received from other scales), $T_{KBP}>0$, as black solid; and
      magnetic energy removed by compression,
      $T_{KBP}<0$ as red dotted.} The error level of the transfer analysis is
    $2.5\cdot10^2\,$erg cm$^{-2}$ s$^{-1}$ (see Appendix
    \ref{SEC:TRANSAPP}).}
  \label{FIG:RUNC_LIN}
\end{figure}

With the sources of magnetic energy quantified on a scale-by-scale
basis, we can now identify the source of the small-scale magnetic
energy seen in Figure \ref{FIG:LINGROWTH}.  This energy is seen to
peak at scales between $60$ and $120\,$km.  As we can see in Figure
\ref{FIG:RUNC_LIN}, the ``cascade'' dominates the production of
magnetic energy only at scales below $20\,$km.  Just as for
incompressible turbulent dynamos \citep{MAP05}, there exists a range
of scales where the amplification of the magnetic field is dominated
by injection {of energy} from turbulent \neu{stretching} while the magnetic
cascade dominates \neu{only} at smaller scales. The source of the
$\sim100\,$km-scale magnetic energy is a dynamo process and not the
turbulent tangling of larger-scale field lines.  As the magnetic
energy is predominantly produced at a scale of {$65\,$km} due to
the stretching by fluid motions which operate predominately at a scale
of {$140\,$km} {(the peak of $-T_{BKT}$ in Figure
  \ref{FIG:RUNC_LIN}),} this suggests that the field lines stretched
by these motions have a scale between {$45$ and $120\,$km.}  This
is due to the fact that all transfers occur between three fields whose
wave-vectors must form a triad (see Figure
\ref{FIG:TRIAD}).\footnote{{See Equation (\ref{eq:convotriad}), describing
the production of
    magnetic energy at wavenumber $k$.  The expression
    involves the transform of the product of the magnetic field (at
    some scale, $p$) and the velocity field (at another scale, $q$).
    By the convolution theorem, we see that $\vec{p}$ must be given by
    $\vec{k}-\vec{q}$.}}  This allows us to rule out
Alfv\'enic \neu{turbulent induction,} i.e., small-scale velocity fluctuations interacting
with a large-scale field to produce small-scale magnetic energy, as an
important source of the small-scale magnetic energy in the
simulations.  Furthermore, as the three scales involved in the dynamo
all lie in the inertial range, we can identify the \murama dynamo as a
turbulent small-scale dynamo.

%  \label{FIG:TRIAD}
\begin{figure}[htbp]
  \plottwo{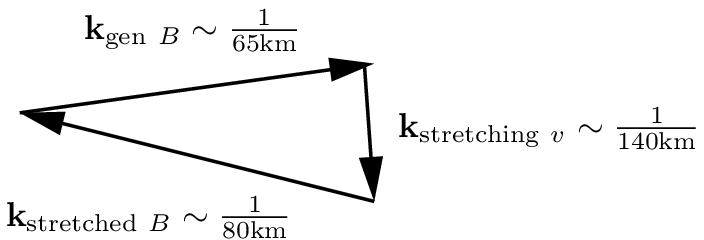}{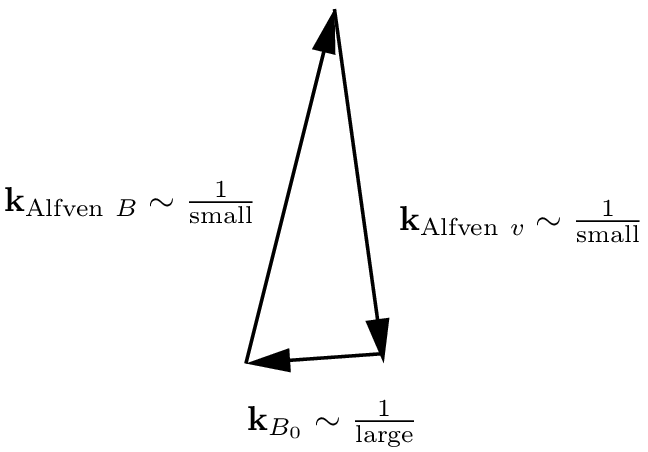}
  \caption{Illustration of triadic transfers.  Left: Case for the
    linear phase of the \murama dynamo \runultraL.  Fluid motions predominately
    at a scale of $\sim140\,$km create magnetic energy predominately
    at a scale of $\sim65\,$km.  As the three wave-vectors must form
    a triad, the scale of the magnetic field being stretched must have
    a scale of $80\pm40\,$km.  Right: The Alfv\'enic case:
    small-scale velocity fluctuations interact with a large-scale
    field to produce small-scale magnetic energy.}
  \label{FIG:TRIAD}
\end{figure}

For a small-scale dynamo, the smallest eddies provide the fastest
stretching and this should be reflected in the transfer analysis.  We
would expect both the dominant scale of stretching motions and of
magnetic field production to decrease as the Reynolds numbers are
increased.  That is, we expect the dynamo action to move to smaller
and smaller scales.  This is, in fact, what we find in the analysis.
The dominant scale of stretching motions is {$200\,$km for \rune,
$180\,$km for \runc, and $140\,$km for \runultraLa while the dominant
scales of magnetic field production are $110\,$km, $100\,$km, and
$65\,$km.}  This provides further evidence that the \murama dynamo is a
turbulent {small-scale} dynamo.

\section{Nonlinear/saturated phase}
\label{SEC:NONLINEAR}

{The analysis of the nonlinear phase of the dynamo is quite
  similar to that already presented for the linear phase.  For the
  sake of brevity, we describe only the differences between the two
  phases without figures.  The magnetic energy spectra peak at scales
  slightly larger than the kinetic Taylor scales, $\lambda_K$.  For
  scales larger than $\lambda_K$, a power-law $k_\perp^\beta$ with
  $\beta=0.4\pm0.1$ can be fit.  This is not as
  steep as in the linear regime and does not appear to be sensitive to
  the Reynolds number.  \neu{It might be sensitive, however, to the large-scale flow
geometry which will be investigated in a future work.}  There is a net transfer of energy, at {2\% for \runea (4\% for \runca and 8\% for \runultraL)}
  of the total magnetic energy generation rate, to fluid motions {\sl
    via} the magnetic tension force for scales smaller than $50\,$km.
  These are short-wavelength Alfv\'en waves.  Magnetosonic waves are
  also produced, at {0.5\% for \runea (1\% for \runca and 2\% for \runultraL)} of the magnetic energy generation rate,
  for scales smaller than $100\,$km.  The saturation of the
  large-scale magnetic energy is reflected by a near balance of
  generation by stretching and losses to the compressive ``cascade,''
  $T_{KBT}(k) + T_{KBP}(k) \approx 0$ for wavenumbers corresponding to
  scales above a few hundred km.  We also note that the representative
  scale at which fluid motions are doing work against the magnetic tension
  force has increased, for \runc, from the linear phase {(where it is
    $180\,$km) up to $250\,$km.}  This result is similar to the result
  for an incompressible small-scale dynamo where in the saturated
  phase (as opposed to the linear phase), forcing-scale eddies
  dominate the energy injection as compared to turbulent eddies
  \citep{AMP05a}.  The representative scale for the production of
  magnetic energy {in \runca is $130\,$km in the saturated state.  From
    triadic considerations, we expect that predominately magnetic
    field with scales between $85$ and $270\,$km is stretched.}  All
  results are again indicative of a small-scale turbulent dynamo as
  all scales in the triad are much smaller than the $\approx1\,$Mm
  energy-containing scale of the granulation convection.}

\section{\neu{$P_M<1$}}
\label{SEC:PRANDTL}

\neu{While our dynamo calculations include much of the physics
  believed to be present in the solar photosphere,} \AD{the limited
  computational resources available today do not accommodate} \neu{realistic
  values of the magnetic and kinetic Reynolds numbers.  Moreover, the
  solar value of their ratio, the magnetic Prandtl number
  $\sim10^{-5}$, will remain unachievable for strongly stratified,
  radiative small-scale dynamo simulations for many years to come.  We
  can, however, take the first steps in this direction by employing
  the grid size of \runultraLa with the magnetic diffusivity of \runc.
  This run, \runprandtla (initialized from 15 minutes into \runultraLa
  to reduce the computational expense) has $P_M\sim0.8$ and an
  exponential growth rate of $\gamma\approx1/1200\,$s$^{-1}$ (see Table
  \ref{TABLE:RUNS} and Figure \ref{FIG:EMVST}).  The lower growth
  rate compared to \runca is likely due to a decreased
  ratio of $Re_M/Re_M^C$ ($Re_M^C$ increases with decreasing $P_M$,
  e.g., \citealt{ScHaBr+2005}). The} \AD{results of the} \neu{transfer analysis are very
  similar to those shown in Figure \ref{FIG:RUNC_LIN}.  We identify
  the dominant scale of stretching motions as $150\,$km (close to that
  of \runultraL) while the dominant scales of magnetic field
  production is $85\,$km (intermediary to  \runca and \runultraL).
  This indicates that the magnetic field being stretched has a scale
  of $130\pm70\,$km.  We conclude that the small-scale turbulent
  dynamo action is essentially unchanged} \AD{in comparison to the runs with $P_M>1$. }

\section{Summary and Conclusions}

We have determined that the \murama surface dynamo is a turbulent
small-scale dynamo.  The dynamo growth rate, $\gamma$, increases with
Reynolds number, {consistent with the picture of dynamo action
  moving to smaller scales.  At scales larger than the peak of
  magnetic energy, the} magnetic spectrum is a power-law indicating a
self-similar (e.g., turbulent) dynamo mechanism.  The magnetic energy
spectrum peaks at scales an order of magnitude smaller than the
energy-containing scale of fluid motions (the granulation scale).
This peak moves to smaller scales with increasing $Re_M$.  {By
  deriving spectral transfer analysis for anisotropic,
  compressible MHD, we are able to identify the source of this
  small-scale magnetic energy.}  The analysis ruled out two
{other possible} mechanisms for the generation of the small-scale
magnetic field: the tangling of larger-scale magnetic field lines associated with
the turbulent cascade and Alfv\'enization of small-scale velocity
fluctuations (or ``turbulent induction'').  We demonstrated that the
source of the magnetic field was due, {rather,} to stretching
motions in the inertial range of the turbulence.  The inertial motions
stretch predominantly small-scale magnetic field to produce more
small-scale magnetic field.  All three scales are significantly
smaller than the granulation scale and move to even smaller scales
with increasing Reynolds numbers.  This positively identifies the
dynamo process as a small-scale turbulent dynamo.

{We also identified the key differences between the \murama
  small-scale dynamo and those studied in the isotropic,
  incompressible MHD case.  The presence of compressive motions opens
  up a new mechanism for the transfer of kinetic to magnetic energy. Namely,
  compression becomes involved in the ``cascade'' of magnetic energy
  to smaller scales, resulting in a net increase (summed over all
  scales) of magnetic energy.  This mechanism accounts for only
  5\% of the rate of magnetic energy generated (and slightly decreasing
  with increasing $Re_M$).  It is
  accomplished by Mm-scale fluid motions compressing larger scale
  magnetic structures (which are generated by stretching motions) into
  scales much smaller ($\la20\,$km) than the peak in the magnetic
  energy spectrum.  This magnetic energy is then, presumably,
  dissipated by the magnetic diffusivity present in the simulations.}

We {have} quantified the scale-dependent anisotropy of the fluid
motions in our convection simulations via measurements of the
second-order {structure functions.}  {The flow is anisotropic at
  scales larger than $50\,$km and this appears to play a role in the
  large-scale magnetic energy spectrum generated by the dynamo in that
  it differs from the isotropic Kazantsev result and varies with the
  degree of anisotropy of the flow.} The slope is significantly less
steep than the $k^{+3/2}$ {law} of the Kazantsev incompressible
small-scale dynamo, {but steepens with decreasing anisotropy of
  the flow as measured by} the inertial-range velocity increments.
This anisotropy decreases with Reynolds number as stronger horizontal
fluctuations are generated by the turbulence (strong vertical
fluctuations being induced by the convective driving).

Our analysis {suggests} that {solar} surface convection is capable, via
a turbulent small-scale dynamo, of generating and sustaining
small-scale magnetic field.  Of course, dynamo action {in the Sun} will also be
present {in} lower layers where more kinetic energy is available and
where stratification and rotation {may also} lead to an inverse
cascade to large scales ($\alpha-$effect).  These two types of
turbulent dynamos are likely intimately related.  Even without a
large-scale seed field {local small-scale} dynamo action should occur
(e.g., during solar grand minima and for {slowly rotating}
stars).  In our simulation no large-scale background field is present.
Turbulent induction of such a field (from a {global} dynamo)
{is an additional} mechanism for the generation of small-scale
field: \neu{it might obscure the presence of a small-scale dynamo.
On the other hand, the large-scale magnetic field and the small-scale
field produced from it will be subject to Ohmic decay:} % Depending on the dissipation time-scale of this co-existing
%dynamo's magnetic field,  might be
%obscured.  Or,
 small-scale dynamo action might sustain magnetic field
that would otherwise be lost. % to \neu{dissipation.} %Ohmic decay.
  A future study should
quantify at what background field strength the small-scale dynamo no
longer dominates the production of small-scale magnetic field {(as
was done by \citealt{CaEmWe2003} for the Boussinesq case).}

\neu{The $P_M\ll1$ case will remain inaccessible} \AD{for a long
 time to come} \neu{for ``realistic'' small-scale dynamo simulations of the solar
  surface.
%Solar and stellar plasma have $P_M\ll1$ while our simulations were
%conducted for $P_M\ga1$.
Nonetheless,} \AD{evidence of the existence of $P_M\ll1$ dynamos for idealistic simulations has
now been given 
  \citep{PMM+05}.}  \neu{When possible, radiative MHD
magneto-convection calculations for $P_M\la0.1$ should be carried out to determine if the magnetic field
generated has a different morphology for this case.  Differences in the
  magnetic structuring has been demonstrated for idealistic
  simulations of small-scale dynamo action with $P_M\la0.1$
  \citep{IsScCo+2007,ScIsCo+2007}, but these differences likely only
  affect the solar magnetic field at sub-kilometer scales.}

\neu{Small-scale dynamo action is not suppressed for $P_M\ll1$ in
  idealistic simulations \citep{PMM+05,IsScCo+2007,ScIsCo+2007}.  We
  have shown that ``realistic'' simulations of dynamo action in
  strongly stratified, radiative MHD with partial ionization and
  little recirculation for $P_M\ga1$ (first achieved by
  \citealt{VoSc2007}) also do not suppress the small-scale dynamo nor
  do they supplant it with another mechanism.  It is reasonable, then,
  to infer that small-scale dynamo action can occur for $P_M\ll1$
  combined with solar-like stratification, radiation, ionization, and
  recirculation.}

{\bf Acknowledgments}

The authors would like to acknowledge fruitful discussions with
A. Fournier, P. Mininni, W.-C. M\"uller, and D. Rosenberg.  This work
has been supported by the Max-Planck Society in the framework of the
Interinstitutional Research Initiative ``Turbulent transport and ion
heating, reconnection and electron acceleration in solar and fusion
plasmas'' of the MPI for Solar System Research Lindau and the
Institute for Plasma Physics, Garching (project MIF-IF-A-AERO8047).

%%%%%%%%%%%%%%%%%%%%%%%%%%%%%%%%%%%%%%%%%%%%%%%%%%%%%%%%%%%%%%%%

\appendix

\section{Appendix}

\subsection{Derivation of transfer functions}
\label{SEC:TRANSFER}

Transfer analysis for the incompressible case is well-known (see
\citealt{Kr1967} or for a more recent exposition, e.g.,
\citealt{AMP05a}).  Here, we generalize the theory to the compressible
and strongly stratified case.  The latter requires a departure from
the usual isotropic (and periodic) Fourier basis.  Given a complete
orthonormal basis, $\phi_k(\vec{x})$, and a function $g(\vec{x})$,
\begin{equation}
g(\vec{x}) = \sum_k \hat{g}(k) \phi_k(\vec{x})
\end{equation}
with
\begin{equation}
\hat{g}(k) \equiv \int_\Omega g(\vec{x})\phi_k(\vec{x})dx^3
\end{equation}
where $\Omega$ is the analysis volume.
Integrating the spatial density, $g(\vec{x})$, of a global quantity, $G$, over the volume allows us to identify $\hat{g}(k)$ as a ``spectral density'' of $G$ in $k$ space,
\begin{equation}
G \equiv \int_\Omega g(\vec{x}) dx^3 =\int_\Omega \sum_k \hat{g}(k) \phi_k(\vec{x})dx^3
=  \sum_k \hat{g}(k) \int_\Omega \phi_k(\vec{x})dx^3 = \sum_k \hat{g}(k)\,.
\end{equation}
For example, the magnetic spectral energy density {satisfies}
\begin{equation}
E_M \equiv \int_\Omega e_M dx^3 = \sum_k E_M(k)\,,
\end{equation}
where $e_M \equiv\frac{1}{8\pi}|\vec{B}|^2$ is the magnetic energy density (in Gaussian units) and $E_M$ is the total magnetic energy.
From Parseval's theorem we see that
\begin{equation}
\int_\Omega\vec{B}(\vec{x})\cdot\vec{B}(\vec{x})dx^3=
\sum_k \hat{\vec{B}}(k)\cdot\hat{\vec{B}}^{\ast}(k)
\end{equation}
where $\cdot^{\ast}$ specifies conjugate (depending on the basis).  This allows
us to identify that the magnetic spectral energy density is
\begin{equation}
E_M(k) \equiv \frac{1}{8\pi}\hat{\vec{B}}(k)\cdot\hat{\vec{B}}^{\ast}(k)\,.
\end{equation}
We can derive its temporal evolution {(the magnetic component of
  the energy balance equation)} by projecting the induction equation,
\begin{eqnarray}
\partial_t\vec{B} +\vec{v}\cdot\nabla\vec{B}= \vec{B}\cdot\nabla\vec{v} - \vec{B}\nabla\cdot\vec{v} + \eta \nabla^2 \vec{B} \resp{+\vec{v}\nabla\cdot\vec{B}}\,,
\label{eq:induction}
\end{eqnarray}
onto the basis functions and then taking the dot product of this expression with
$\frac{1}{4\pi}\hat{\vec{B}}^{\ast}(k)$. We add the conjugate of the result and
divide the sum by two to derive that the time evolution of the
spectral magnetic energy density,
\begin{equation}
\frac{d}{dt} E_M(k) = {T_{KB}(k)+T_{IB}(k)}\,,
\end{equation}
is given by the magnetic energy transfer functions
(representing the transfer of magnetic energy in $k-$space). In general, the
transfer functions, $T_{XY}(k)$, measure the rate of energy transfer
from energy reservoir $X$ to the $k-$component of reservoir $Y$ (also the
work done on field $Y$ by field $X$ at wavenumber $k$).  For
$T_{XY}(k) >0$ ($<0$) energy is received (lost) by the $k-$component
of $Y$.
{In particular, $T_{KB}(k)\equiv T_{KBT}(k)+T_{KBP}(k)$} denotes the energy transfer rate from the kinetic energy reservoir to the {$k-$component
of the} magnetic energy reservoir
through {the dynamo term (stretching against the magnetic tension force),}\footnote{{Abbreviated expressions are used where the conjugate is assumed.  The full expression is ${T_{KBT}(k)} = \frac{1}{8\pi}\left(\hat{\vec{B}}(k)\cdot\widehat{\left[\vec{B}\cdot\nabla\vec{v}\right]}^{\ast}(k)+\hat{\vec{B}}^{\ast}(k)\cdot\widehat{\left[\vec{B}\cdot\nabla\vec{v}\right]}(k)\right)$.}} 
\begin{equation}
{T_{KBT}(k)} = \frac{1}{4\pi}\hat{\vec{B}}(k)\cdot\widehat{\left[\vec{B}\cdot\nabla\vec{v}\right]}^{\ast}(k)\,,
\label{eq:convotriad}
\end{equation}
and {compression against magnetic pressure,}
\begin{equation} 
{T_{KBP}(k) = - \frac{1}{4\pi}\hat{\vec{B}}(k)\cdot\widehat{\left[\vec{B}\nabla\cdot\vec{v}\right]}^{\ast}(k)-\frac{1}{4\pi}\hat{\vec{B}}(k)\cdot\widehat{\left[\vec{v}\cdot\nabla\vec{B}\right]}^{\ast}(k)}\,.
\label{eq:TKBP}
\end{equation}
$T_{IB}(k)$ represents the (negative of) energy loss rate to Joule heating,
\begin{equation}
T_{IB}(k) =\frac{\eta}{4\pi}\hat{\vec{B}}(k)\cdot\widehat{\left[ \nabla^2\vec{B}\right]}^{\ast}(k)\,.
\end{equation}
\resp{$\sum_kT_{IB}(k) =\frac{\eta}{4\pi} \int_\Omega\vec{B}\cdot(\nabla^2\vec{B})dx^3$.  $T_{MM}(k)$ signifies magnetic energy generated by the numerical
artifact of magnetic monopoles,
\begin{equation}
T_{MM}(k) = \frac{1}{4\pi}\hat{\vec{B}}(k)\cdot\widehat{\left[\vec{v}\nabla\cdot\vec{B}\right]}^{\ast}(k)\,.
\end{equation}
}

The kinetic spectral energy density can be expressed
\begin{equation}
E_K(k) = \frac{1}{4}\left( \hat{\vec{v}}(k)\cdot\widehat{\left[\rho\vec{v}\right]}^{\ast}(k)
+\widehat{\left[\rho\vec{v}\right]}(k)\cdot\hat{\vec{v}}^{\ast}(k)
\right)\,.
\end{equation}
Note that this is different from the velocity spectrum,
\begin{equation}
E_V(k) = \frac{1}{2}\hat{\vec{v}}(k)\cdot\hat{\vec{v}}^{\ast}(k)\,,
\end{equation}
in the incompressible theory of Kolmogorov (see, e.g., \citealt{F95}).
The time evolution of the kinetic spectral energy density is given by
\begin{equation}
\frac{d}{dt}E_K(k) =
\frac{1}{2}\hat{\vec{v}}(k)\cdot\partial_t\widehat{\left[\rho\vec{v}\right]}^{\ast}(k) +
\frac{1}{2}\widehat{\left[\rho\vec{v}\right]}^{\ast}(k)\cdot\partial_t\hat{\vec{v}}(k)\,.
\end{equation}
\neu{This is an alternative expression to the one derived by \citet{MiKi1995} for the non-magnetic case.}
The kinetic energy transfer functions can be derived similarly to the expression for the magnetic transfer
utilizing the conservation of momentum density equation, (see, e.g., \citealt{VSS+05}),
\begin{eqnarray}
\partial_t(\rho \vec{v}) + \nabla\cdot(\rho\vec{v}\vec{v}) = -\nabla P + \rho \vec{F} +
\mu\nabla\cdot\uuline{\tau} + \frac{1}{c} (\vec{J}\times\vec{B})\,,
\label{eq:momentum}
\end{eqnarray}
and the conservation of mass density, $\rho$,
\begin{eqnarray}
\partial_t\rho + \nabla\cdot(\rho \vec{v}) = 0\,,
\label{eq:mass}
\end{eqnarray}
which can be combined to form the momentum equation in terms of the velocity
\begin{eqnarray}
\partial_t\vec{v} + \vec{v}\cdot\nabla\vec{v} = -\frac{1}{\rho}\nabla P + \vec{F} +
\frac{\mu}{\rho}\nabla\cdot\uuline{\tau} + \frac{1}{\rho c} (\vec{J}\times\vec{B})\,.
\label{eq:momentumV}
\end{eqnarray}
Taking the appropriate dot products and integrating over the total
volume, we {derive that the kinetic part of the energy balance
  equation,}
\begin{equation}
{\frac{d}{dt} E_K(k)  = \mathfrak{f}(k) + T_{KK}(k)+ T_{BK}(k)+ T_{IK}(k)}
\label{eq:dEVspec}
\end{equation}
is given by the energy injection into the volume by body forces
\begin{equation}
{\mathfrak{f}(k) = \frac{1}{2}\hat{\vec{v}}(k)\cdot\widehat{\left[\rho\vec{F}\right]}^{\ast}(k)
+\frac{1}{2}\widehat{\left[\rho\vec{v}\right]}^{\ast}(k)\cdot\hat{\vec{F}}(k)}
\end{equation}
and the sum of the kinetic energy transfer functions.
$T_{KK}(k) = T_{KKA}(k)+T_{KKC}(k)$ is the transfer of kinetic energy inside the kinetic energy reservoir (``cascade''), by {advection}
\begin{equation}
T_{KKA}(k)  = -\frac{1}{2}\hat{\vec{v}}(k)\cdot\widehat{\left[\vec{v}\cdot\nabla(\rho\vec{v})\right]}^{\ast}(k)
-\frac{1}{2}\widehat{\left[\rho\vec{v}\right]}^{\ast}(k)\cdot\widehat{\left[\vec{v}\cdot\nabla\vec{v}\right]}(k)\,,
\end{equation}
\resp{($\sum_k T_{KKA}(k)  = -\int_\Omega \vec{v} \cdot \nabla e_K dx^3$)} and compressible (irrotational) motions
\begin{equation}
T_{KKC}(k)  = -\frac{1}{2}\hat{\vec{v}}(k)\cdot\widehat{\left[\rho\vec{v}\nabla\cdot\vec{v}\right]}^{\ast}(k)\,.
\end{equation}
\resp{($\sum_k T_{KKC}(k)  = -\int_\Omega e_K \nabla\cdot\vec{v} dx^3$).}
$T_{IK}(k)= T_{IKC}(k)+T_{IKD}(k)$ represents energy transferred from the internal energy reservoir to
the kinetic energy reservoir
by compression,
\begin{equation}
T_{IKC}(k)  = - \frac{1}{2}\hat{\vec{v}}(k)\cdot\widehat{\left[\nabla P\right]}^{\ast}(k)
-\frac{1}{2}\widehat{\left[\rho\vec{v}\right]}^{\ast}(k)\cdot\widehat{\left[\frac{1}{\rho}\nabla P\right]}(k)\,,
\end{equation}
and viscous dissipation,
\begin{equation}
T_{IKD}(k)  =  - \frac{1}{2}\hat{\vec{v}}(k)\cdot\widehat{\left[\mu\nabla\cdot\uuline\tau\right]}^{\ast}(k)
-\frac{1}{2}\widehat{\left[\rho\vec{v}\right]}^{\ast}(k)\cdot\widehat{\left[\frac{\mu}{\rho}\nabla\cdot\uuline\tau\right]}(k)\,.
\end{equation}
{$T_{BK}(k)=T_{BKT}(k)+T_{BKP}(k)$} represents energy transferred
from the magnetic energy reservoir to the kinetic energy reservoir by
the work of the Lorenz force {both via magnetic tension,
\begin{equation}
T_{BKT}(k)  =  \frac{1}{8\pi}\hat{\vec{v}}(k)\cdot\widehat{\left[\vec{B}\cdot\nabla\vec{B}\right]}^{\ast}(k)
+\frac{1}{8\pi}\widehat{\left[\rho\vec{v}\right]}^{\ast}(k)\cdot\widehat{\left[\frac{1}{\rho}\vec{B}\cdot\nabla\vec{B}\right]}(k)
\end{equation}
and via magnetic pressure,
\begin{equation}
T_{BKP}(k)  =  \frac{1}{8\pi}\hat{\vec{v}}(k)\cdot\widehat{\left[-\frac{1}{2}\nabla|\vec{B}|^2\right]}^{\ast}(k)
+\frac{1}{8\pi}\widehat{\left[\rho\vec{v}\right]}^{\ast}(k)\cdot\widehat{\left[-\frac{1}{2\rho}\nabla|\vec{B}|^2\right]}(k)\,.
\label{eq:TBKP}
\end{equation}}

\neu{For an ideal gas with adiabatic index, $\gamma$, the internal
  energy density is given by $e_I = P/(\gamma-1)$ and the pressure is
  given by $P = \rho\mathfrak{R}T$ where $T$ is the temperature and
  $\mathfrak{R}$ the gas constant.
The internal spectral energy density can then be expressed
\begin{equation}
E_I(k) = \frac{\mathfrak{R}}{2(\gamma-1)}\left( \hat{T}(k)\hat{\rho}^{\ast}(k)
+\hat{\rho}(k)\hat{T}^{\ast}(k)
\right)\,.
\end{equation}
The time evolution of the internal spectral energy density is then
\begin{equation}
\frac{d}{dt}E_I(k) =
\frac{\mathfrak{R}}{\gamma-1}\hat{T}(k)\partial_t\hat{\rho}^{\ast}(k) +
\frac{\mathfrak{R}}{\gamma-1}\hat{\rho}^{\ast}(k)\partial_t\hat{T}(k)\,.
\end{equation}
Using the conservation of mass, Equation (\ref{eq:mass}),
and the energy equation for temperature,
\begin{equation}
\rho\frac{\mathfrak{R}}{\gamma-1}\left(\partial_tT+\vec{v}\cdot\nabla T\right) = \nabla\cdot(\kappa\nabla T)
- P \nabla\cdot\vec{v} +Q_{rad}+\mu(\uuline{\tau}\cdot\nabla)^\top\vec{v}\nonumber\\
+ (\frac{c}{4\pi})^2\frac{1}{\sigma}|\vec{J}|^2\,,
\label{eq:temp}
\end{equation}
we find the that time evolution of the internal spectral energy density,}
\begin{equation}
\frac{d}{dt} E_I(k)  = \mathfrak{f}_I(k) + T_{I}(k)\,,
\label{eq:dEIspec}
\end{equation}
is given by the radiative heating (cooling), $\sum_k\mathfrak{f}_I(k)
= \int_\Omega Q_{rad}dx^3$, and the sum of the internal energy
transfer function, $T_I(k)$.  $T_I(k) = T_{II}(k) + T_{KI}(k) +
T_{BI}(k)$.  $T_{II}(k)$ is the transfer of internal energy inside the
internal energy \neu{reservoir: $T_{II}(k) = -\frac{\mathfrak{R}}{\gamma-1}
\hat{T}(k)\widehat{\left[\nabla\cdot(\rho\vec{v})\right]}^\ast(k)
-\frac{\mathfrak{R}}{\gamma-1}\hat{\rho}^\ast(k)
\widehat{\left[\vec{v}\cdot\nabla T\right]}(k)
+\hat{\rho}^\ast(k)\widehat{\left[\frac{1}{\rho}\nabla\cdot(\kappa\nabla T)\right]}(k)$}.
\neu{$T_{KI}(k)=T_{KIC}(k)+T_{KID}(k)$ represents transfer of kinetic energy to internal
  energy by work against pressure gradients, $T_{KIC}(k) =
-\hat{\rho}^\ast(k)\widehat{\left[\frac{1}{\rho}P\nabla\cdot\vec{v}\right]}(k)$ and viscous
  dissipation, $T_{KID}(k) =
\hat{\rho}^\ast(k)\widehat{\left[\frac{1}{\rho}\mu(\uuline{\tau}\cdot\nabla)^\top\vec{v}\right]}(k)$.}  $T_{BI}(k)$ represents transfer of magnetic energy to
internal energy by Joule heating, \neu{$T_{BI}(k) =\hat{\rho}^\ast(k)\widehat{\left[\frac{1}{\rho}
  (\frac{c}{4\pi})^2\frac{1}{\sigma}|\vec{J}|^2\right]}(k)$} where
\begin{equation}
\sum_kT_{IB}(k) + T_{BI}(k) =-(\frac{c}{4\pi})\frac{1}{\sigma} \int_\Omega\nabla\cdot\left(\vec{J}\times\vec{B}\right)dx^3\,.
\end{equation}
As this is the integral over a divergence term, we recognize that the
magnetic energy lost and the Joule heat gained differ only by a
surface term.
{The sum of transfers between kinetic and magnetic energies is}
\begin{equation}
\sum_kT_{KB}(k) + T_{BK}(k)\resp{+ T_{MM}(k)} =\frac{1}{4\pi} \int_\Omega\nabla\cdot\left((\vec{v}\times\vec{B})\times\vec{B}\right)dx^3{\equiv\mathfrak{P}_{\partial\Omega}\,,}
\label{eq:Pflux1}
\end{equation}
{where $\mathfrak{P}_{\partial\Omega}$ is the {\sl inductive component} of the Poynting flux into the domain.  From this,}
we see that the time derivative of the total magnetic energy,
\begin{eqnarray}
\frac{d}{dt} E_M = \sum_k  T_{B}(k)   = \mathfrak{p}_{\partial\Omega}   - \sum_k T_{BK}(k)- \sum_k T_{BI}(k) \,,
\label{eq:dEMspec}
\end{eqnarray}
is given by the inflow of electromagnetic energy (negative Poynting flux) at the boundary, $\partial\Omega$,
\begin{equation}
\mathfrak{p}_{\partial\Omega} = -\frac{c}{4\pi}\int_\Omega \nabla\cdot \left( \vec{E} \times \vec{B}\right) dx^3
 = \int_{\partial\Omega} -\frac{c}{4\pi}\left( \vec{E} \times \vec{B}\right) \cdot d\vec{A}\,,
\end{equation}
the Lorentz force work against internal fluid motions, $\sum_k
T_{BK}(k)$, and diffusive losses to Joule heating, $\sum_k T_{BI}(k)$.

{The right hand side of Equation (\ref{eq:Pflux1}) is equivalent
  to a surface term via Gauss's divergence theorem,
\begin{equation} 
\mathfrak{P}_{\partial\Omega}\equiv \frac{1}{4\pi}\int_{\partial\Omega}\left((\vec{v}\times\vec{B})\times\vec{B}\right)\cdot d\vec{A}\,.
\end{equation}
This surface term is zero for common choices of boundary conditions, \AD{hence,}
\begin{equation}
\sum_kT_{KB}(k) = -\sum_k T_{BK}(k)\,.
\label{eq:LorentzTransfer}
\end{equation}
\neu{This follows from the fact that $T_{KB}$ and $T_{BK}$ measure the transfer}
of energy between the magnetic and kinetic energy reservoirs.  We can also
show that
\begin{equation}
\sum_kT_{KBP}(k) + T_{BKP}(k) = \frac{1}{4\pi}\int_{\partial\Omega}|\vec{B}|^2\vec{v}\cdot d\vec{A}\,.
\end{equation}
This is another surface term \AD{and}
\begin{equation}
\sum_kT_{KBP}(k) = -\sum_k T_{BKP}(k)
\label{eq:PressureTransfer}
\end{equation}
\neu{as these two functions measure energy transfer via the
interaction of compression and magnetic pressure.}
In incompressible MHD, the second term of $T_{KBP}$ in Equation (\ref{eq:TKBP}),
that is
\begin{equation}
T_{BB}(k) = -\frac{1}{4\pi}\hat{\vec{B}}(k)\cdot\widehat{\left[\vec{v}\cdot\nabla\vec{B}\right]}^{\ast}(k)\,,
\end{equation}
is identified as the transfer rate of magnetic energy to other scales
within the magnetic energy reservoir (the magnetic energy
``cascade'').  As the first term of $T_{KBP}$ is identically zero and
$\sum_k T_{BKP}(k)=0$ for incompressible flow, $\sum_k T_{BB}(k)=0$
(transfer is internal to magnetic energy reservoir).  This is not true
for compressible flow and the magnetic ``cascade'' cannot be separated
from compressive kinetic energy transfers.    From Equations
(\ref{eq:LorentzTransfer}) and (\ref{eq:PressureTransfer}), we also
find
\begin{equation}
\sum_kT_{KBT}(k) = -\sum_k T_{BKT}(k)\,,
\label{eq:TensionTransfer}
\end{equation}
\neu{as} that these two functions measure energy
transfer via the stretching of magnetic field lines against the
magnetic tension force.}  \neu{A similar analysis shows,
$\sum_kT_{IB}(k) = -\sum_k T_{BI}(k)$,
$\sum_kT_{IKD}(k) = -\sum_k T_{KID}(k)$,
and $\sum_kT_{IKC}(k) = -\sum_k T_{KIC}(k)$.}

\subsection{Application to \murama simulations}
\label{SEC:TRANSAPP}

For stratified convection, we use as the orthonormal basis functions a two-dimensional
Fourier basis multiplied by vertical cardinal basis functions,
\begin{equation}
\phi_{k_x,k_y,z'} = \frac{1}{2\pi}e^{-i(k_xx+k_yy)}\delta(z-z')\,.
\end{equation}
To study the generation (or loss of) magnetic energy at a given scale, the
resulting transfer functions are projected onto a one-dimensional wavenumber,
$k_{{\perp}}^2=k_x^2+k_y^2$, and summed over the vertical direction,
\begin{eqnarray}
T_{KB}(k_{{\perp}})\equiv\sum_z\sum_{k_{{\perp}}=\sqrt{k_x^2+k_y^2}} T_{KB}(k_x,k_y,z)\,.
\end{eqnarray}
To quantify the source of magnetic energy generation at the scale
$2\pi/k_\perp$, we need only measure $T_{KBT}$ and $T_{KBP}$.  The
resistive transfer, $T_{IB}$, {mostly} serves as a
sink for magnetic energy at small ($\la5\Delta x$) scales.  To complete
the physical picture, we also measure $T_{BKT}$ and $T_{BKP}$, the {rate} at which
kinetic energy is lost to (gained from) the magnetic field. 
\neu{Transfers between internal and kinetic energies may be ignored  as we are only interested in generation and loss of magnetic energy in a dynamo analysis.} %\footnote{\neu{Additionally, the internal energy transfer derived for an ideal gas does not apply to \murama which uses a tabulated equation of state.}}}
 As shown in
the previous section, Equation (\ref{eq:Pflux1}), the sum of these four transfers
is equal to the inflow of electromagnetic energy from
the {inductive component} of the Poynting flux, $\mathfrak{P}_{\partial\Omega}$, into the domain,
\begin{equation} 
\sum_{z_{bot}<z<z_{top}}\sum_{k_{{\perp}}}T_{KBT}+T_{KBP}+T_{BKT}+T_{BKP}+T_{MM}+\Delta_{chain}=\mathfrak{P}_{\partial\Omega}\,.
\label{eq:Pflux2}
\end{equation}
Here $T_{MM}(k) =
\frac{1}{4\pi}\hat{\vec{B}}(k)\cdot\widehat{\left[\vec{v}\nabla\cdot\vec{B}\right]}^{\ast}(k)$
denotes the influence of the numerical artifacts of magnetic monopoles and
\begin{equation}
\Delta_{chain}\equiv\int_\Omega\vec{v}\cdot\nabla e_M-\frac{1}{4\pi}\vec{B}\cdot\left(\vec{v}\cdot\nabla\vec{B}\right)dx^3
\label{eq:DelChain}
\end{equation}
arises because of numerical inaccuracies in the chain rule, $0.5
dB^2/dx=BdB/dx$.  Multiplying \muram's 5-pt stencil for the derivative
of $B$ by $B$ yields,
\begin{equation}
B \frac{dB}{dx} - \frac{\Delta x^4}{30} B B^{(5)}\,,
\end{equation}
while the numerical derivative of $B^2/2$ yields 
\begin{equation}
B \frac{dB}{dx} - \frac{\Delta x^4}{36} (6 \frac{dB}{dx} B^{(4)}+ 12 \frac{d^2B}{dx^2} B^{(3)}+ 5 B^{(3)} B^{(4)})\,.
\end{equation}
As strong gradients of the magnetic field exist in the downflows,
$o(B) \ll o(B^{(n)})$ and significant departures, {$\approx8\%$
of the total transfer rate,} from the chain rule
occur.  We track this through direct calculation of $\Delta_{chain}$.
Lower order errors occur for the two grid points closest to the top
and bottom boundaries.  For this reason we choose {our analysis
domain such that the boundary, $\partial\Omega$, is}
halfway between the second and third grid points
from the bottom (and top).  $\mathfrak{P}_{\partial\Omega}$ is
interpolated using Newton divided differences and the error in
calculating Equation (\ref{eq:Pflux2}) is computed for each snapshot.
As Equation (\ref{eq:DelChain}) accounts for the chain-rule error only
for the largest and most easily calculated contribution, some small
residual error is found.  This error, {always less than 3\%,} is assumed to be spread equally
over $k_\perp-$space and is reported for all plots
of transfer functions.

%\clearpage
%\bibliographystyle{apj}
%\bibliography{journals,joushort,bigfile}

\begin{thebibliography}{50}
\expandafter\ifx\csname natexlab\endcsname\relax\def\natexlab#1{#1}\fi

\bibitem[{{Alexakis} {et~al.}(2005){Alexakis}, {Mininni}, \&
  {Pouquet}}]{AMP05a}
{Alexakis}, A., {Mininni}, P.~D., \& {Pouquet}, A. 2005, \pre, 72, 046301

\bibitem[{{Batchelor}(1950)}]{Ba1950}
{Batchelor}, G.~K. 1950, Proceedings of the Royal Society of London. Series A,
  Mathematical and Physical Sciences, 201, 405

\bibitem[{{Batchelor}(1953)}]{Ba1970}
---. 1953, {Theory of Homogeneous Turbulence} (Cambridge, England: Cambridge
  University Press)

\bibitem[{{Boldyrev} \& {Cattaneo}(2004)}]{BoCa2004}
{Boldyrev}, S. \& {Cattaneo}, F. 2004, Physical Review Letters, 92, 144501

\bibitem[{{Brun} {et~al.}(2004){Brun}, {Miesch}, \& {Toomre}}]{BMT04}
{Brun}, A.~S., {Miesch}, M.~S., \& {Toomre}, J. 2004, \apj, 614, 1073

\bibitem[{{Cattaneo}(1999)}]{C99}
{Cattaneo}, F. 1999, \apjl, 515, L39

\bibitem[{{Cattaneo} {et~al.}(2003){Cattaneo}, {Emonet}, \&
  {Weiss}}]{CaEmWe2003}
{Cattaneo}, F., {Emonet}, T., \& {Weiss}, N. 2003, \apj, 588, 1183

\bibitem[{{Covas} {et~al.}(1997){Covas}, {Tworkowski}, {Brandenburg}, \&
  {Tavakol}}]{CoTwBr+1997}
{Covas}, E., {Tworkowski}, A., {Brandenburg}, A., \& {Tavakol}, R. 1997, \aap,
  317, 610

\bibitem[{{Debliquy} {et~al.}(2005){Debliquy}, {Verma}, \&
  {Carati}}]{DeVeCa2005}
{Debliquy}, O., {Verma}, M.~K., \& {Carati}, D. 2005, Physics of Plasmas, 12,
  042309

\bibitem[{{Eyink}(2006)}]{Ey2006}
{Eyink}, G.~L. 2006, Journal of Fluid Mechanics, 549, 191

\bibitem[{{Field} \& {Blackman}(2002)}]{FiBl2002}
{Field}, G.~B. \& {Blackman}, E.~G. 2002, \apj, 572, 685

\bibitem[{{Field} {et~al.}(1999){Field}, {Blackman}, \& {Chou}}]{FiBlCh1999}
{Field}, G.~B., {Blackman}, E.~G., \& {Chou}, H. 1999, \apj, 513, 638

\bibitem[{Frisch(1995)}]{F95}
Frisch, U. 1995, {Turbulence, The Legacy of A. N. Kolmogorov} (Cambridge, UK:
  Cambridge University Press)

\bibitem[{{G{\'o}mez} {et~al.}(2005{\natexlab{a}}){G{\'o}mez}, {Mininni}, \&
  {Dmitruk}}]{GMD05b}
{G{\'o}mez}, D.~O., {Mininni}, P.~D., \& {Dmitruk}, P. 2005{\natexlab{a}},
  Advances in Space Research, 35, 899

\bibitem[{{G{\'o}mez} {et~al.}(2005{\natexlab{b}}){G{\'o}mez}, {Mininni}, \&
  {Dmitruk}}]{GMD05}
---. 2005{\natexlab{b}}, Physica Scripta Volume T, 116, 123

\bibitem[{{Hagenaar} {et~al.}(2003){Hagenaar}, {Schrijver}, \&
  {Title}}]{HaScTi2003}
{Hagenaar}, H.~J., {Schrijver}, C.~J., \& {Title}, A.~M. 2003, \apj, 584, 1107

\bibitem[{{Haugen} \& {Brandenburg}(2004)}]{HaBr2004b}
{Haugen}, N.~E.~L. \& {Brandenburg}, A. 2004, \pre, 70, 036408

\bibitem[{{Iskakov} {et~al.}(2007){Iskakov}, {Schekochihin}, {Cowley},
  {McWilliams}, \& {Proctor}}]{IsScCo+2007}
{Iskakov}, A.~B., {Schekochihin}, A.~A., {Cowley}, S.~C., {McWilliams}, J.~C.,
  \& {Proctor}, M.~R.~E. 2007, Physical Review Letters, 98, 208501

\bibitem[{{Kraichnan}(1967)}]{Kr1967}
{Kraichnan}, R.~H. 1967, Physics of Fluids, 10, 1417

\bibitem[{{Kraichnan}(1971)}]{Kr1971}
---. 1971, Journal of Fluid Mechanics, 47, 525

\bibitem[{{Krause} \& {Raedler}(1980)}]{KrRa1980}
{Krause}, F. \& {Raedler}, K.~H. 1980, {Mean-field magnetohydrodynamics and
  dynamo theory}, ed. K.~H. Krause, F. \&~Raedler

\bibitem[{{Maltrud} \& {Vallis}(1993)}]{MaVa1993}
{Maltrud}, M.~E. \& {Vallis}, G.~K. 1993, Physics of Fluids, 5, 1760

\bibitem[{{Meneguzzi} {et~al.}(1981){Meneguzzi}, {Frisch}, \&
  {Pouquet}}]{MeFrPo1981}
{Meneguzzi}, M., {Frisch}, U., \& {Pouquet}, A. 1981, Physical Review Letters,
  47, 1060

\bibitem[{{Mininni} {et~al.}(2005{\natexlab{a}}){Mininni}, {Alexakis}, \&
  {Pouquet}}]{MAP05}
{Mininni}, P., {Alexakis}, A., \& {Pouquet}, A. 2005{\natexlab{a}}, \pre, 72,
  046302

\bibitem[{{Mininni} {et~al.}(2008){Mininni}, {Alexakis}, \&
  {Pouquet}}]{MiAlPo2008}
{Mininni}, P.~D., {Alexakis}, A., \& {Pouquet}, A. 2008, \pre, 77, 036306

\bibitem[{{Mininni} {et~al.}(2005{\natexlab{b}}){Mininni}, {Ponty},
  {Montgomery}, {Pinton}, {Politano}, \& {Pouquet}}]{MPM+05}
{Mininni}, P.~D., {Ponty}, Y., {Montgomery}, D.~C., {Pinton}, J.-F.,
  {Politano}, H., \& {Pouquet}, A. 2005{\natexlab{b}}, \apj, 626, 853

\bibitem[{{Miura} \& {Kida}(1995)}]{MiKi1995}
{Miura}, H. \& {Kida}, S. 1995, Physics of Fluids, 7, 1732

\bibitem[{{Monchaux} {et~al.}(2009){Monchaux}, {Berhanu}, {Auma{\^i}tre},
  {Chiffaudel}, {Daviaud}, {Dubrulle}, {Ravelet}, {Fauve}, {Mordant},
  {P{\'e}tr{\'e}lis}, {Bourgoin}, {Odier}, {Pinton}, {Plihon}, \&
  {Volk}}]{MoBeAu+2009}
{Monchaux}, R., {Berhanu}, M., {Auma{\^i}tre}, S., {Chiffaudel}, A., {Daviaud},
  F., {Dubrulle}, B., {Ravelet}, F., {Fauve}, S., {Mordant}, N.,
  {P{\'e}tr{\'e}lis}, F., {Bourgoin}, M., {Odier}, P., {Pinton}, J., {Plihon},
  N., \& {Volk}, R. 2009, Physics of Fluids, 21, 035108

\bibitem[{{Monchaux} {et~al.}(2007){Monchaux}, {Berhanu}, {Bourgoin}, {Moulin},
  {Odier}, {Pinton}, {Volk}, {Fauve}, {Mordant}, {P{\'e}tr{\'e}lis},
  {Chiffaudel}, {Daviaud}, {Dubrulle}, {Gasquet}, {Mari{\'e}}, \&
  {Ravelet}}]{MoBeBo+2007}
{Monchaux}, R., {Berhanu}, M., {Bourgoin}, M., {Moulin}, M., {Odier}, P.,
  {Pinton}, J.-F., {Volk}, R., {Fauve}, S., {Mordant}, N., {P{\'e}tr{\'e}lis},
  F., {Chiffaudel}, A., {Daviaud}, F., {Dubrulle}, B., {Gasquet}, C.,
  {Mari{\'e}}, L., \& {Ravelet}, F. 2007, Physical Review Letters, 98, 044502

\bibitem[{{Ossendrijver}(2003)}]{Os2003}
{Ossendrijver}, M. 2003, \aapr, 11, 287

\bibitem[{{Petrovay} \& {Szakaly}(1993)}]{PeSz1993}
{Petrovay}, K. \& {Szakaly}, G. 1993, \aap, 274, 543

\bibitem[{{Pietarila Graham} {et~al.}(2009){Pietarila Graham}, {Danilovic}, \&
  {Sch{\"u}ssler}}]{PiGrDaSc2009}
{Pietarila Graham}, J., {Danilovic}, S., \& {Sch{\"u}ssler}, M. 2009, \apj,
  693, 1728

\bibitem[{{Ponty} {et~al.}(2005){Ponty}, {Mininni}, {Montgomery}, {Pinton},
  {Politano}, \& {Pouquet}}]{PMM+05}
{Ponty}, Y., {Mininni}, P.~D., {Montgomery}, D.~C., {Pinton}, J.-F.,
  {Politano}, H., \& {Pouquet}, A. 2005, Physical Review Letters, 94, 164502

\bibitem[{{Ponty} {et~al.}(2007){Ponty}, {Mininni}, {Pinton}, {Politano}, \&
  {Pouquet}}]{PoMiPi+2007}
{Ponty}, Y., {Mininni}, P.~D., {Pinton}, J.-F., {Politano}, H., \& {Pouquet},
  A. 2007, New Journal of Physics, 9, 296

\bibitem[{{Pouquet} {et~al.}(1976){Pouquet}, {Frisch}, \& {Leorat}}]{PFL76}
{Pouquet}, A., {Frisch}, U., \& {Leorat}, J. 1976, Journal of Fluid Mechanics,
  77, 321

\bibitem[{{Rogachevskii} \& {Kleeorin}(1997)}]{RoKl1997}
{Rogachevskii}, I. \& {Kleeorin}, N. 1997, \pre, 56, 417

\bibitem[{{S{\'a}nchez Almeida}(2003)}]{SaAl2003}
{S{\'a}nchez Almeida}, J. 2003, \aap, 411, 615

\bibitem[{{S{\'a}nchez Almeida}(2004)}]{SA04}
{S{\'a}nchez Almeida}, J. 2004, in ASP Conf. Ser. 325: The Solar-B Mission and
  the Forefront of Solar Physics, ed. T.~{Sakurai} \& T.~{Sekii}, 115--+

\bibitem[{{S{\'a}nchez Almeida} {et~al.}(2003){S{\'a}nchez Almeida}, {Emonet},
  \& {Cattaneo}}]{SaAlEmCa2003b}
{S{\'a}nchez Almeida}, J., {Emonet}, T., \& {Cattaneo}, F. 2003, in
  Astronomical Society of the Pacific Conference Series, Vol. 307, Astronomical
  Society of the Pacific Conference Series, ed. J.~{Trujillo-Bueno} \&
  J.~{Sanchez Almeida}, 293--+

\bibitem[{{Schekochihin} {et~al.}(2004){Schekochihin}, {Cowley}, {Taylor},
  {Maron}, \& {McWilliams}}]{ScCoTa+2004}
{Schekochihin}, A.~A., {Cowley}, S.~C., {Taylor}, S.~F., {Maron}, J.~L., \&
  {McWilliams}, J.~C. 2004, \apj, 612, 276

\bibitem[{{Schekochihin} {et~al.}(2005){Schekochihin}, {Haugen}, {Brandenburg},
  {Cowley}, {Maron}, \& {McWilliams}}]{ScHaBr+2005}
{Schekochihin}, A.~A., {Haugen}, N.~E.~L., {Brandenburg}, A., {Cowley}, S.~C.,
  {Maron}, J.~L., \& {McWilliams}, J.~C. 2005, \apjl, 625, L115

\bibitem[{{Schekochihin} {et~al.}(2007){Schekochihin}, {Iskakov}, {Cowley},
  {McWilliams}, {Proctor}, \& {Yousef}}]{ScIsCo+2007}
{Schekochihin}, A.~A., {Iskakov}, A.~B., {Cowley}, S.~C., {McWilliams}, J.~C.,
  {Proctor}, M.~R.~E., \& {Yousef}, T.~A. 2007, New Journal of Physics, 9, 300

\bibitem[{{Stein} {et~al.}(2003){Stein}, {Bercik}, \& {Nordlund}}]{StBeNo2003}
{Stein}, R.~F., {Bercik}, D., \& {Nordlund}, {\AA}. 2003, in Astronomical
  Society of the Pacific Conference Series, Vol. 286, Current Theoretical
  Models and Future High Resolution Solar Observations: Preparing for ATST, ed.
  A.~A. {Pevtsov} \& H.~{Uitenbroek}, 121--+

\bibitem[{{Title} \& {Schrijver}(1998)}]{TiSc1998}
{Title}, A.~M. \& {Schrijver}, C.~J. 1998, in Astronomical Society of the
  Pacific Conference Series, Vol. 154, Cool Stars, Stellar Systems, and the
  Sun, ed. R.~A. {Donahue} \& J.~A. {Bookbinder}, 345--+

\bibitem[{{Trujillo Bueno} {et~al.}(2004){Trujillo Bueno}, {Shchukina}, \&
  {Asensio Ramos}}]{TrBuShAsRa2004}
{Trujillo Bueno}, J., {Shchukina}, N., \& {Asensio Ramos}, A. 2004, \nat, 430,
  326

\bibitem[{{Verma} {et~al.}(2005){Verma}, {Ayyer}, \& {Chandra}}]{VeAyCh2005}
{Verma}, M.~K., {Ayyer}, A., \& {Chandra}, A.~V. 2005, Physics of Plasmas, 12,
  082307

\bibitem[{{V{\"o}gler}(2003)}]{V03}
{V{\"o}gler}, A. 2003, PhD thesis, G{\"o}ttingen University

\bibitem[{{V{\"o}gler} \& {Sch{\"u}ssler}(2007)}]{VoSc2007}
{V{\"o}gler}, A. \& {Sch{\"u}ssler}, M. 2007, \aap, 465, L43

\bibitem[{{V{\"o}gler} {et~al.}(2005){V{\"o}gler}, {Shelyag}, {Sch{\"u}ssler},
  {Cattaneo}, {Emonet}, \& {Linde}}]{VSS+05}
{V{\"o}gler}, A., {Shelyag}, S., {Sch{\"u}ssler}, M., {Cattaneo}, F., {Emonet},
  T., \& {Linde}, T. 2005, \aap, 429, 335

\bibitem[{{Weygand} {et~al.}(2007){Weygand}, {Matthaeus}, {Dasso}, {Kivelson},
  \& {Walker}}]{WeMaDa+2007}
{Weygand}, J.~M., {Matthaeus}, W.~H., {Dasso}, S., {Kivelson}, M.~G., \&
  {Walker}, R.~J. 2007, Journal of Geophysical Research (Space Physics), 112,
  10201

\bibitem[{{Zhou}(1993)}]{Zh1993}
{Zhou}, Y. 1993, Physics of Fluids, 5, 1092

\end{thebibliography}

%\clearpage

%% the file, uncomment the \plotone, \plottwo, and
%% If you need a layout that cannot be achieved with \plotone or
%% \plottwo, you can invoke the  package directly with the
%% \includegraphics command or use \plotfiddle. For more information,
%% please see the tutorial on "Using Electronic Art with " in the
%% documentation section at the  Web site,
%% http://www.journals.uchicago.edu/AAS/AASTeX.
%%

%    \label{FIG:EMVST}
% \label{FIG:SCALES}

%\begin{figure}[htbp]
%  \plotone{fig_struct}
%  \caption{Second-order structure functions  for velocities, $S_2^v(l)$, for the {subsurface part
%    (lower $800\,$km) of the computational} box, of \runultraLa
%    versus length, $l$, for the vertical direction: $v_z$ (plus signs)
%    and horizontal directions: $v_x$ (dotted) and $v_y$ (dashed).  The
%    ratios demonstrate an anisotropy in the velocity distributions at
%    scales larger than $\approx50\,$km.  The gray line indicates a
%    power-law fit of $l^{-(\beta+1)}$ of $\beta=-1.65\pm0.06$ for the
%    vertical direction. }
%  \label{FIG:STRUCT}
%\end{figure}

%  \label{FIG:GAMMA}
%  \label{FIG:LINGROWTH}

%  \label{FIG:TRANSFERS}

%  \label{FIG:RUNC_LIN}

%  \label{FIG:TRIAD}

%\clearpage
%\label{TABLE:RUNS}

\end{document}